\documentclass[showpacs,aps,preprintnumbers,prb,amsmath,amssymb,twocolumn,floatfix,superscriptaddress]{revtex4-2}
\usepackage{graphicx}
\usepackage{hyperref}
\usepackage{color}
\usepackage{bm}
\usepackage{amssymb}
\usepackage{amsmath}
\usepackage[normalem]{ulem}
\usepackage[T1]{fontenc}
\usepackage{multirow}
\usepackage{xcolor}

\newcommand{\lc}{\lowercase}
\begin{document}
\title{Off-stoichiometric effect on magnetic and electron transport properties of Fe$_2$VAl$_{1.35}$ in respect to Ni$_2$VAl; Comparative study}
\author{A.~\'{S}lebarski}
\affiliation{Institute of Low Temperature and Structure Research, Polish Academy of Sciences, Ok\'{o}lna 2, 50-422 Wroc{\l}aw, Poland}
\affiliation{Centre for Advanced Materials and Smart Structures, 
Polish Academy of Sciences, Ok\'{o}lna 2, 50-422 Wroc\l aw, Poland}
\affiliation{$^{\star}$ Author to whom correspondence should be addressed: andrzej.slebarski@us.edu.pl}
\author{M. Fija\l kowski}
\affiliation{Institute of Physics,
University of Silesia in Katowice, 75 Pu\l ku Piechoty 1, 41-500 Chorz\'ow, Poland}
\affiliation{Centre for Advanced Materials and Smart Structures, 
Polish Academy of Sciences, Ok\'{o}lna 2, 50-422 Wroc\l aw, Poland}
\author{J.~Deniszczyk} 
\affiliation{Institute of Materials Engineering, University of Silesia in Katowice, 75 Pu{\l}ku Piechoty 1A, \mbox{41-500 Chorz\'{o}w}, Poland}
\author{M.~M.~Ma\'ska} 
\affiliation{Institute of Theoretical Physics, Wroclaw University of Science and Technology, Wybrze\.{z}e Wyspia\'{n}skiego 27, 50-370 Wroc\l aw, Poland}
\author{D.~Kaczorowski}
\affiliation{Institute of Low Temperature and Structure Research, Polish Academy of Sciences, Ok\'{o}lna 2, 50-422 Wroc{\l}aw, Poland}
\affiliation{Centre for Advanced Materials and Smart Structures, 
Polish Academy of Sciences, Ok\'{o}lna 2, 50-422 Wroc\l aw, Poland}

\begin{abstract}

Density functional theory (DFT) calculations confirm that the structurally ordered Fe$_2$VAl Heusler alloy is nonmagnetic  narrow-gap semiconductor. This compound is apt to form various disordered modifications with high concentration of antisite defects. We study the effect of structural disorder on the electronic structure, magnetic, and electronic transport properties of the full Heusler alloy Fe$_2$VAl and its off-stoichiometric equivalent Fe$_2$VAl$_{1.35}$. 
Data analysis in relation to {\it ab initio} calculations indicates an appearance of antisite disorder mainly due to Fe--V and Fe--Al stoichiometric variations. The data for weakly magnetic Fe$_2$VAl$_{1.35}$ are discussed in respect to  Ni$_2$VAl.
Fe$_2$VAl$_{1.35}$ can be classified as a nearly ferromagnetic metal with a pronounced spin glassy contribution, which, however, does not give a predominant effect on its thermoelectric properties. The figure of merit $ZT$ is at 300 K about 0.05 for the Fe sample and 0.02 for Ni one, respectively.
However, it is documented that the narrow $d$ band resulting from Fe/V site exchange  can be responsible for the unusual temperature dependencies of the physical properties of the Fe2TiAl$_{1.35}$ alloy, characteristic of strongly correlated electron systems.
As an example, the magnetic susceptibility of Fe$_2$VAl$_{1.35}$ exhibits singularity characteristic of a Griffiths phase, appearing as an inhomogeneous electronic state below $T_G\sim 200$ K. We also performed numerical analysis which supports the Griffiths phase scenario.

\end{abstract}


\maketitle
\section{Introduction}

Cubic Heusler compounds, known as Heusler alloys, constitute a large family of materials, exhibiting a variety of interesting properties, both with respect to basic and applied investigations \cite{Felser2016,Chatterjee2022}. 
In particular, in the last two decades, it has been experimentally demonstrated that some Heusler alloys can exhibit superconductivity \cite{Klimczuk2012} as well as topological effects \cite{Lin2020}, they are also promising materials for thermoelectric applications \cite{Bourgault2023}. These alloys continue to be an active area of research in condensed matter physics.
Particular attention is paid to examining the impact of a widely understood atomic disorder on the physical properties of these alloys.
As an example, for a number of Fe-based Heusler alloys, disorder caused by dopants, off-stoichiometry of the system, or the presence of antisite (AS) Fe defects has been shown to enhance their thermoelectric properties \cite{Bourgault2023}, as well as it may give a reason for the appearance of exotic phenomena related to magnetic instabilities, which are in many cases associated with the proximity of a quantum critical point (QCP) \cite{Sato2010,Slebarski2009}. However, the origin of these behaviors is still controversial. A good example of such quantum phenomena seems to be Fe$_2$VAl. The investigations of the  off-stoichiometric Fe$_{2+x}$V$_{1-x}$Al and  Fe$_{2}$VAl$_{1-x}$ equivalents have suggested the presence of a metal-insulator transition resulting in observed singularities and enhancements of thermodynamic quantities near the expected ferromagnetic QCP \cite{Naka2012,Naka2016}.
Fe$_2$VAl is a nonmagnetic and nonmetallic (semimetallic) material, exhibiting a narrow pseudogap at the Fermi level \cite{Weht1998,Singh1998,Guo1998}. Graf {\it et al.} \cite{Graf2011} reported that Fe$_2$VAl is located at a nonmagnetic node on the Slater-Pauling curve of the spontaneous ferromagnetic moment $m$ in
multiples of Bohr magnetons $\mu_B$, where $m$  scales with the
total number of valence electrons following the rule $m =Z-24$, and $Z$ is the number of valence electrons (see also \cite{Galanakis2002}).
However, the weak ferromagnetism of this compound can be activated by atomic defects of AS Fe, as documented by DFT calculations.
Numerous previous reports \cite{Nishino2001,Matsushita2002,Slebarski2006} 
have documented that antisite defects associated with Fe and V sites create local heterogeneous electronic states in Fe$_2$VAl, quite different compared to the nonmagnetic and semimetallic state of a defect free sample. 
The {\it ab initio} band structure calculation gives a magnetic AS Fe at the V site, when it is surrounded by four Fe atoms occupying Fe Wyckoff positions. 
There are also other AS defects that can be formed in Fe$_2$VAl$_{1+\delta}$, we therefore expect a variety of emergent phenomena resulting from the disorder introduced as a result of antisite defects and off-stoichiometry. 
We document experimentally complex magnetic behavior for Fe$_2$VAl$_{1.35}$. Our attention will be focused on the low temperature enhancement in magnetic susceptibility with singularity $\chi\sim T^{-1+\lambda}$, as well as magnetization  $M\sim B^{\lambda}$ behavior, both are  reminiscent of a Griffiths-McCoy singularity \cite{Griffiths1969,comment2}.

The experimental data for Fe$_2$VAl$_{1.35}$ are discussed with respect to those of nonmagnetic and metallic Ni$_2$VAl~\cite{Rocha1999,Wen2017,Wang2020}, which recently has been reported as a candidate for superconductivity \cite{Sreenivasa2016}. Our investigations have  not supported Ni$_2$VAl as a superconductor. 
It has been shown, however, that disorder of AS-type generates a weak magnetic moment located on Ni at AS position. In consequence of the AS disorder, the resistivity has a  $\rho \sim -\ln T$ behavior in the temperatures $T<10$ K, which characterizes Ni$_2$VAl as a diluted Kondo system. Moreover,  accompanying spin fluctuations are detected with characteristic maximum in $\chi(T)$ at about 120 K,  which seems to be interesting (see Sec. III.A and D).

\section{Experimental and computational details}
\subsection{Measurements}
\begin{figure}[h!]
\includegraphics[width=0.48\textwidth]{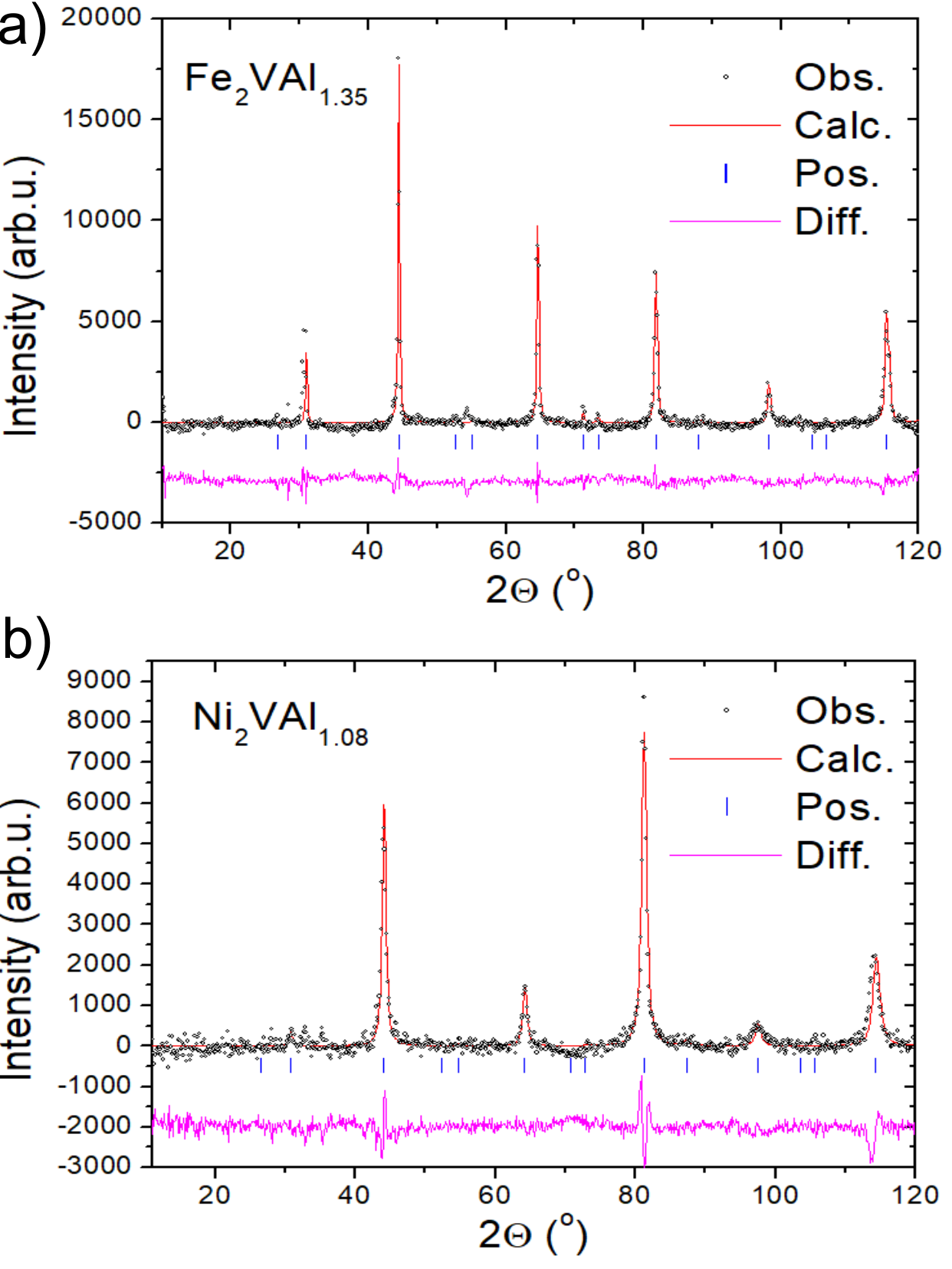}
\caption{\label{fig:F+N_XRD}
Plot of Rietveld refinement for Fe$_2$VAl$_{1.35}$ (a) and Ni$_2$VAl$_{1.08}$ (b). Black dots - observed pattern, red line - calculated, blue ticks - Bragg peaks positions, magenta line - the difference. The Ni$_2$VAl$_{1.08}$ alloy is extremely hard, so the stresses when preparing the sample for XRD analysis cause a greater spread of background intensities in the XRD diffraction pattern.
}
\end{figure}

Polycrystalline samples of  Fe$_2$VAl$_{1.35}$, V$_2$FeAl and Ni$_2$VAl$_{1.08}$ were prepared by the arc melting technique and subsequent annealing  at 800$^{o}$C for 2 weeks. The products were examined by x-ray diffraction (XRD) analysis (PANalytical Empyrean diffractometer equipped with a Cu K$\alpha_{1,2}$ source) and found to have a face-centered cubic L2$_1$ crystal structure (space group F$m\bar{3}m$). The XRD patterns were analyzed with the Rietveld refinement method using the Fullprof Suite set of programs \cite{Rodriguez1993}. Figure \ref{fig:F+N_XRD} shows an XRD pattern for Fe$_2$VAl$_{1.35}$ (a) and Ni$_2$VAl$_{1.08}$ (b) with Rietveld refinements. 
The results presented in Table \ref{tab:TableXRD} were obtained for each sample with the weighted-profile $R$ factors~\cite{Toby2006}  $R_{wp}<1.8$\% and $R_{Bragg}<0.9$\%.
\begin{table}[h!]
\caption{Lattice parameters $a$ and stoichiometry.
}
\label{tab:TableXRD}
\begin{tabular}{ccc}
\hline\hline
compound & $a$ (\AA) & composition (at\%) \\
\hline
Fe$_2$VAl$_{1.35}$ &  5.7659(8) & $45.86:23.02:31.13$\\        
Ni$_2$VAl$_{1.08}$ & 5.7996(9) & $49.48:24.25:26.27$\\   
V$_2$FeAl          &  5.9469(5) & $49.55:25.11:25.34$\\ 
\hline\hline
\end{tabular} 
\end{table}
Stoichiometry and homogeneity were checked using an electron energy dispersive spectroscopy (EDS) technique. 
The atomic percentage of the specific element content in  Fe$_2$VAl, Ni$_2$VAl, and V$_2$FeAl is listed in Table \ref{tab:TableXRD}. For Ni$_2$VAl, and V$_2$FeAl it deviates from the nominal composition $2:1:1$ at an acceptable level, while Fe$_2$VAl$_{1.35}$ was identified as off-stoichiometric with excess of Al and with a homogeneous distribution of atoms.

The ac magnetic susceptibility was measured in the temperature range 2-300 K with an ac field of 2 Oe and freqency from 100 Hz to 4 kHz using a Quantum Design PPMS platform. The dc magnetic measurementswere carried out  in the temperature interval $1.7 - 400$ K and magnetic fields up to 7 T employing a Quantum Design superconducting quantum interference device (SQUID) magnetometer. Time-dependent remnant magnetization and high-temperature dc magnetic susceptibility ($300<T<800$ K) were measured using the PPMS platform equipped with a vibrating sample magnetometer (VSM) option. Electrical resistivity and heat capacity measuremments were performed in the temperature range $1.8-300$ K and in external magnetic fields up to 9 T  using the same PPMS platform.

The x-ray photoelectron spectroscopy (XPS) spectra were obtained at room temperature with monochromatized Al $K_{\alpha}$ radiation using a PHI 5700/600 ESCA spectrometer. To obtain good quality XPS spectra, the samples were cleaved and measured in the vacuum of $6\times 10^{-10}$ Torr.

\subsection{Computational methods}
The electronic and magnetic properties of Fe$_2$VAl and Ni$_2$VAl, as well as the off-stoichiometry components, were theoretically studied using the {\em ab initio}, DFT-based full potential linearized augmented plane waves (FP-LAPW) method complemented with local orbitals (LO) \cite{Singh}. The calculations were performed using the WIEN2k (ver. 19.1) package \cite{Blaha2001}.
The atomic core states were treated within the fully relativistic DFT approach. For the local orbitals and valence states, (assumed as follows: V~-~[$3s^23p^6$]$_{\rm LO}$\{$3d^34s^2$\}$_{\rm VB}$; Fe~-~[$3s^23p^6$]$_{\rm LO}$\{$3d^64s^2$\}$_{\rm VB}$; Ni~-~[$3s^23p^6$]$_{\rm LO}$\{$3d^84s^2$\}$_{\rm VB}$ and Al~-~[$2s^22p^6$]$_{\rm LO}$\{$3s^23p^1$\}$_{\rm VB}$) the scalar-relativistic Kohn-Sham formalism was applied with spin-orbit coupling (SOC) accounted for through the second variational method \cite{Singh}. The generalized gradient approximation (GGA) for the exchange-correlation (XC) energy functional was applied in the form derived for solids by Perdew {\em et al.} (PBEsol) \cite{Perdev1996}. For the correlated $d$ states, the XC potential was corrected by on-site Hubbard-like interaction $U$ following the Anisimov {\em at al.} approach \cite{Anisimov1991,Anisimov1993}. In the calculations presented, we assumed the effective Hubbard parameter $U_{\rm eff}$ ($=U-J$) for $d$-states of Fe, Ni, and V we assumed equal to 3~eV.

For simulations of off-stoichiometric systems with antisite atoms, we employed the supercell spanned by doubled primitive vectors of the underlying L2$_1$ primitive cell, comprising eight formula units of Fe$_2$VAl (Ni$_2$VAl), based on which the superstructures were prepared with Al, Fe, Ni, and V atoms located at antisite positions. The calculations were performed for
a basic Fe$_2$VAl  and Ni$_2$VAl structures and superstructures with compositions: Fe$_{16}$(V$_7$Fe$_1$)Al$_8$, Fe$_{16}$(V$_7$Al$_1$)Al$_8$, Fe$_{16}$(V$_5$Al$_3$)Al$_8$, (Fe$_{15}$Al$_1$)V$_8$Al$_8$, (Fe$_{13}$Al$_3$)V$_8$Al$_8$, and  Ni$_{16}$(V$_7$Ni$_1$)Al$_8$. Structural analysis revealed that the antisite atoms in the vanadium sublattice do not change the space group Fm$\bar{3}$m (no. 225) of the original Heusler structure, while those located in the Fe sublattice reduce the space group of the corresponding superstructure to the F4$\bar{3}$m (no. 216). Nevertheless, in all cases, the disorder caused by AS atoms, connected to the anisotropy introduced by the spin-orbit coupling, split the Wyckoff positions of the Heusler structure into several subgroups (see Table \ref{tab:TableDFT}). 

In the presented approach, the parameters decisive for the accuracy of the calculations employing the WIEN2k code, the number of $\vec k$ vectors in the Brillouin zone (BZ), and the plane wave cut-off energy ($K_{\rm max}$) were tested against the total energy convergence. A satisfactory energy precision of few meV for the base Fe$_2$VAl and Ni$_2$VAl was reached with  $12\times12\times12$ k-mesh (163 $\vec k$ vectors in irreducible BZ) and $K_{\rm max}=9/R_{MT}$. The radii of the muffin-tin spheres $R_{MT}$ of 0.1058~nm were assumed as common for all atomic species. These settings were also adopted in the calculations for superstructures.

\section{Magnetic and transport properties in disordered F\lc{e}$_2$VA\lc{l}$_{1.35}$ in reference to spin-fluctuator N\lc{i}$_2$VA\lc{l}, experimental details and discussion}

\subsection{Magnetic properties\label{sec:magn_prop}}
The magnetic and transport properties of Heusler alloys are highly dependent on stoichiometry, as well as the level of atomic disorder. Here we present the magnetic properties of Fe$_2$VAl$_{1+\delta}$ with excess of Al ($\delta\sim 0.35$) with respect to Ni$_2$VAl$_{1+\delta}$ ($\delta\sim 0.08$), we also discuss the impact of antisite defects on the localization of $3d$ electronic states of Fe and Ni in both alloys.
A detailed analysis of the complex magnetic behaviors documented for Fe$_2$VAl$_{1+\delta}$ is also based on other thermodynamic and electron transport studies (in Sec. III.B-D), as well as {\it ab initio} electronic structure calculations, presented in Sec. IV.

\begin{figure}[h!]
\includegraphics[width=0.45\textwidth]{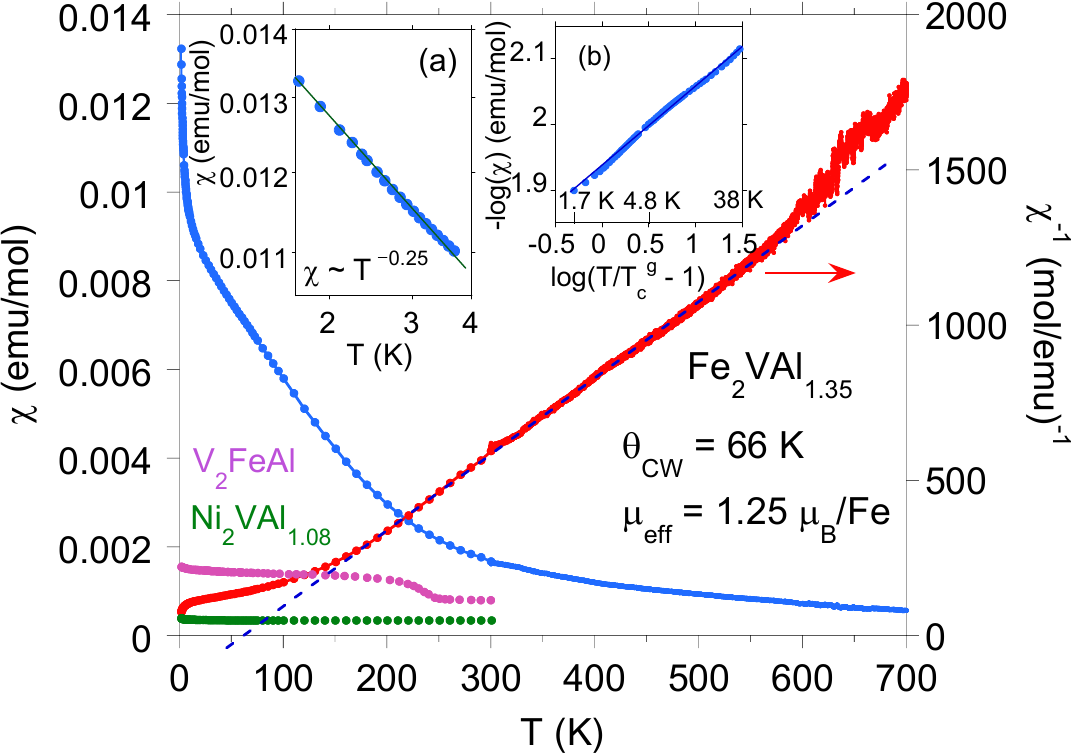}
\caption{\label{fig:CHI_dc_Fe-Ni_inset}
Temperature dependence of dc $\chi$ (blue points) and $1/\chi$ (red points) at 1000 G for Fe$_2$VAl$_{1.35}$. For comparison, the dc susceptibility $\chi(T)$ at 1000 G are shown for $V_2$FeAl (pink points) and for Ni$_2$VAl$_{1.08}$ (green points). Inset (a) shows the divergent behavior of $\chi$ for $T<4$ K, inset (b) presents  $1/\chi$ vs. reduced $T$, $(T/T_C^g -1)$ in the log--log scale.
We attribute the change in the susceptibility of Fe$_2$VAl$_{1.35}$ that occurs at $\sim 580$ K to a diffusion
process that results in a more highly ordered sample above this temperature.
}
\end{figure}

Shown in Fig. \ref{fig:CHI_dc_Fe-Ni_inset} are the dc magnetic susceptibility $\chi$ data plotted as $\chi$ and $1/\chi$ vs. $T$ between 1.7 and 700 K for Fe$_2$VAl$_{1.35}$.  
The susceptibilities of Ni$_2$VAl$_{1.08}$ and V$_2$FeAl measured in the temperature region $T<300$ K are also displayed for comparison.
The $\chi^{-1}(T)$ follows a Curie law above $\sim 600$ K with the effective magnetic moment $\mu_{\rm eff} = 1.26$ $\mu_B$ per one Fe atom in the formula unit. 
A very crude analysis predicts that statistically only one iron
atom contributes to the value of $\mu_{\rm eff}$ per unit cell of Fe$_2$VAl$_{1+\delta}$, 
assuming that only Fe ions contribute to the effective magnetic moment and  $\mu_{\rm eff}$(Fe$^{2+})=5.4$  $\mu_B$. Magnetization as well as the specific heat data suggest one order of magnitude smaller number of magnetic Fe defects in this system (will be discussed).
DFT calculations confirm that this is an Fe ion in the antisite position, while the remaining Fe atoms in the surrounding of the AS defect are {\it non-magnetic} (will be discussed in Sec. IV).
A Curie-Weiss (CW) law is obeyed in the range of $200<T<580$ K, indicating a peculiar magnetic state with random magnetic interactions below $\sim 200$ K and signaling the onset of weak ferromagnetism. The best fit to $\chi \sim (T-\theta_{CW})^{-1}$ gives the CW temperature $\theta_{CW}=66$ K and $\mu_{\rm eff} = 1.25$ $\mu_B$ per Fe atom, i.e., almost four times smaller value of $\mu_{\rm eff}$ than that predicted for Fe$^{2+}$.  The magnetic susceptibility anomaly below 200 K has been found to arise from a distribution of magnetic defects in the sample (cf. Refs. \cite{Slebarski2006}). Similar anomalous behaviors in $\chi$ appear to be characteristic of the family of Heusler alloys containing the magnetic transition metal $M$, regardless of the stoichiometry of the system (cf. $\chi(T)$ data for V$_2$FeAl in Fig. \ref{fig:CHI_dc_Fe-Ni_inset}), while it is not present in {\it almost paramagnetic} Ni$_2$VAl$_{1.08}$.
As an example, a well ordered V$_2$FeAl is expected to be a Pauli paramagnet \cite{comment1,Smith2023}, while a weak magnetization below $\sim 230$ K can be induced in this material by wrong-site iron atoms as a result of incomplete structural ordering, as shown in Fig.~\ref{fig:CHI_dc_Fe-Ni_inset}.
Our investigations suggest a similarity of this weakly magnetic state with short-range magnetic correlations with the behavior of the Griffiths phase (GP) scenario \cite{Griffiths1969}.
The distinct similarities to the Griffiths phase have already been suggested earlier for the off-stoichiometric  Fe$_2$VAl \cite{Naka2016}, as well as for the (FeNi)TiSn alloy \cite{Chatterjee2022a} and disordered Fe$_2$VAl \cite{Slebarski2011} due to the presence of AS defects.   

\begin{figure}[h!]
\includegraphics[width=0.45\textwidth]{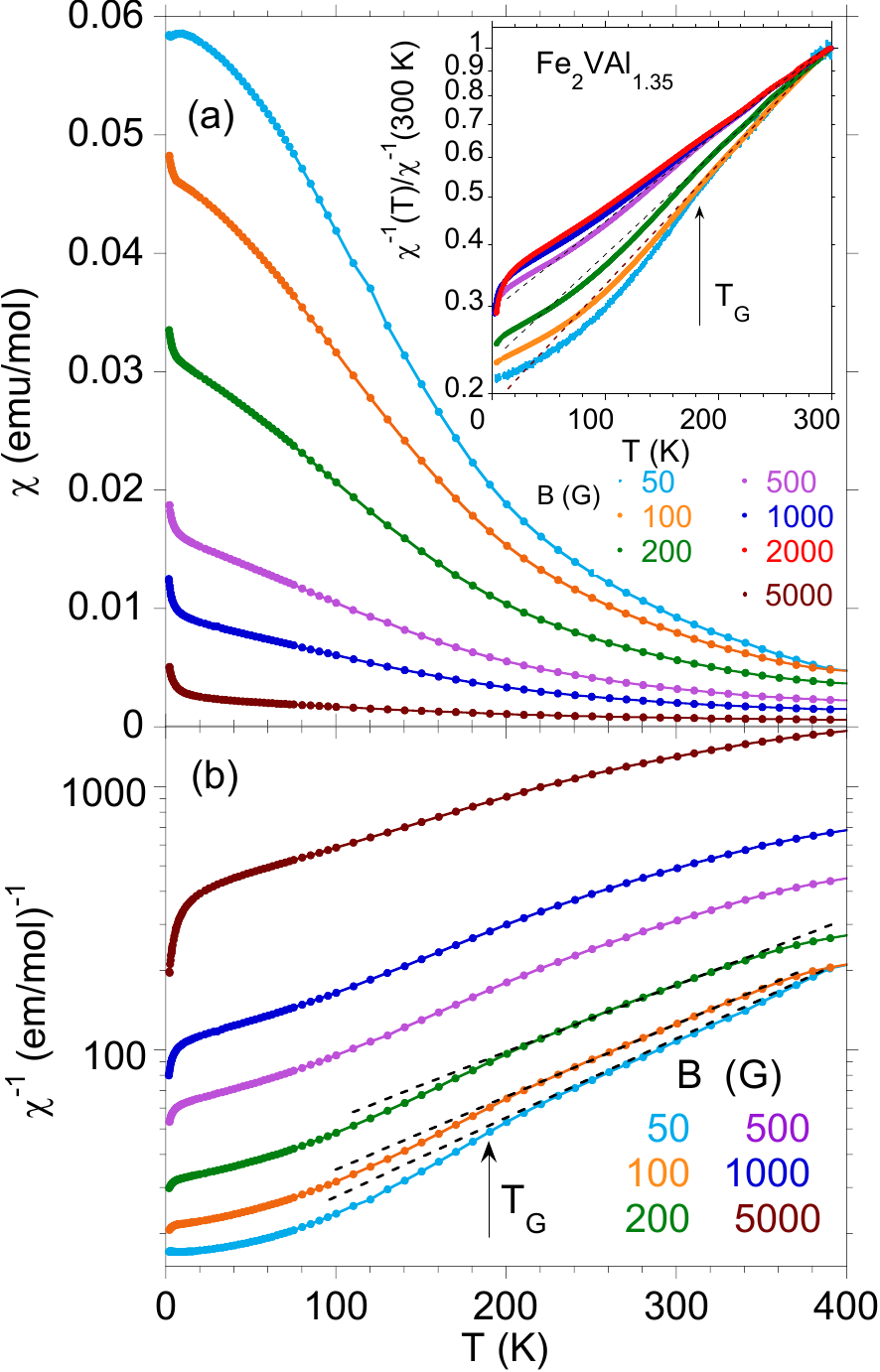}
\caption{\label{fig:Fig_CHI_Fe_sum_inset}
(a) Susceptibility and (b)
inverse susceptibility (in log scale)  vs temperature data from QD SQUID magnetometer,  measured for Fe$_2$VAl$_{1.35}$ in different values of applied fields. The inset shows similar data of $\chi(T)^{-1}$ normalized to $1/\chi$ at 300 K (in log scale) obtained from a commercial QD PPMS-VSM platform. The dashed lines approximate the CW behavior  to low temperatures to show the  {\it downward} effect in $\chi^{-1}$ below temperature $T_G\sim 200$ K. 
}
\end{figure} 

The Griffiths phase consists of magnetic clusters in a
paramagnetic phase much above $T_C$ and forms as a result of the competition between the Kondo effect and the Ruderman-Kittel-Kasuya-Yosida (RKKY) interaction in the presence of disorder \cite{Griffiths1969}. 
Namely, in the temperature region $T_C^g<T<T_G$, where $T_G$ is a Griffiths temperature, the system is considered to exist in the GP that exhibits neither pure paramagnetic behavior nor long-range ferromagnetic (FM) order. 
In this framework, the Griffiths phase is a peculiar state that is predicted to occur in randomly diluted Ising FM systems \cite{Bray1987,Shankar1987}, in which magnetization fails to become an analytic function of the magnetic field over a temperature range $T_C^g<T<T_G$.
Usually, Griffiths singularity is signed by a nonlinear variation of the inverse magnetic susceptibility in the paramagnetic phase \cite{Randeria1985}, namely $\chi^{-1}\propto (T-T_C^g)^{(1-\lambda)}$ ($0<\lambda <1$) \cite{Castro1998,Pramanik2010}, where $T_C^g$ is the critical temperature of random ferromagnetism of the sample where susceptibility tends to diverge.
According to the $\chi$-data shown in Fig. \ref{fig:CHI_dc_Fe-Ni_inset},
the deviation from the CW law is evidenced for $T<200$ K, while below $T=4$ K the dc susceptibility exhibits a power-law behavior, $\chi\sim T^{-n}$, with exponent $n=1-\lambda=0.25$ \cite{Castro1998}. Moreover, for $1.7<T<40$ K $\chi$ can be well characterized by expression $\chi(T)\sim (T-T_C^g)^{-(1-\lambda)}$ with the fitting parameters $T_C^g=1.3$ K and $\lambda=0.7$.

According to the GF scenario, $\chi$ shown in Fig. \ref{fig:CHI_dc_Fe-Ni_inset} deviates from the CW law below $T\equiv T_G=200$ K, while for $T< 4$ K susceptibility exhibits a power-law behavior, $\chi\sim T^{-n}$, with exponent $n=1-\lambda=0.25$ \cite{Castro1998}. Moreover, for $1.7<T<40$ K $\chi$ can be well characterized by expression $\chi(T)\sim (T-T_C^g)^{-(1-\lambda)}$ with the fitting parameters $T_C^g=1.3$ K and $\lambda=0.7$. 
Fig. \ref{fig:Fig_CHI_Fe_sum_inset} shows the $\chi(T)$ dc data measured at various magnetic fields $50<B<5000$ G. 
In the field of 50 G $\chi$ exhibits a maximum at $\sim 9$ K indicative of a magnetic glassy behavior, while $\chi$ measured at larger fields shows divergent behavior at the lowest temperatures.
The inset to Fig. \ref{fig:Fig_CHI_Fe_sum_inset}  presents the inverse susceptibility divided by the value of $\chi^{-1}$ at 300 K in different fields as a function of $T$ from the VSM experiment, to show more details. 
For $T>200$ K $\chi^{-1}$ varies linearly with $T$, following the CW behavior.  However, with the decrease in $T$, a clear downturn in $\chi^{-1}$ is observed at $T\approx 200$ K (much above $T_C^g$) for the measurements performed in dc fields $B\leq 500$ G, indicating non-analytic behavior of $M$ arising from the Griffiths singularity.
The softening of the {\it downward} behavior in $\chi^{-1}$ and the progressive increase of $\chi^{-1}$ in the field (cf. Fig. \ref{fig:Fig_CHI_Fe_sum_inset}) are characteristic properties of the GP state (cf. \cite{Pramanik2010}), both allowed to distinguish the Griffiths singularity from smeared phase transition between the paramagnetic and ferromagnetic states. 

We also comment on the field-induced divergence of the value of $\chi$, shown for Fe$_2$VAl$_{1.35}$ in Fig. \ref{fig:Fig_CHI_Fe_sum_inset}. 
This field-dependent $\chi(T)$ plots may result from a trace amount of magnetic Fe impurities, can be caused by spin fluctuations quenched by the field (cf. \cite{Nishino1997,Tsujii2019}), and/or may be caused by the various size of clusters field-dependent. 
For the first scenario, the appearance of Fe impurities should cause an increase in $\chi$ with an increasing field; this is not a case.
The effect of spin fluctuations on the value of $\chi$  seems to be possible  for Fe$_2$VAl$_{1.35}$ since similar field-induced $\chi(T)$ behavior has also been observed for spin-fluctuator Ni$_2$VAl.
Note that spin fluctuation gives a dominant effect around $T_C$, but is also important much above $T_C$, since the energy scale of spin fluctuations is usually two orders of magnitude larger than $T_C$ for itinerant electron ferromagnets \cite{Moriya1985,Takahashi2013}.
The magnetic properties shown in Fig. \ref{fig:Fig_CHI_Fe_sum_inset} are more likely the result of contributions from both fluctuating moments and cluster size effects, as has been reported for a variety of nanoparticles (e.g., Refs.~\cite{Cox1994,Costro1997,Jo2006}). 

\begin{figure}[h!]
\includegraphics[width=0.45\textwidth]{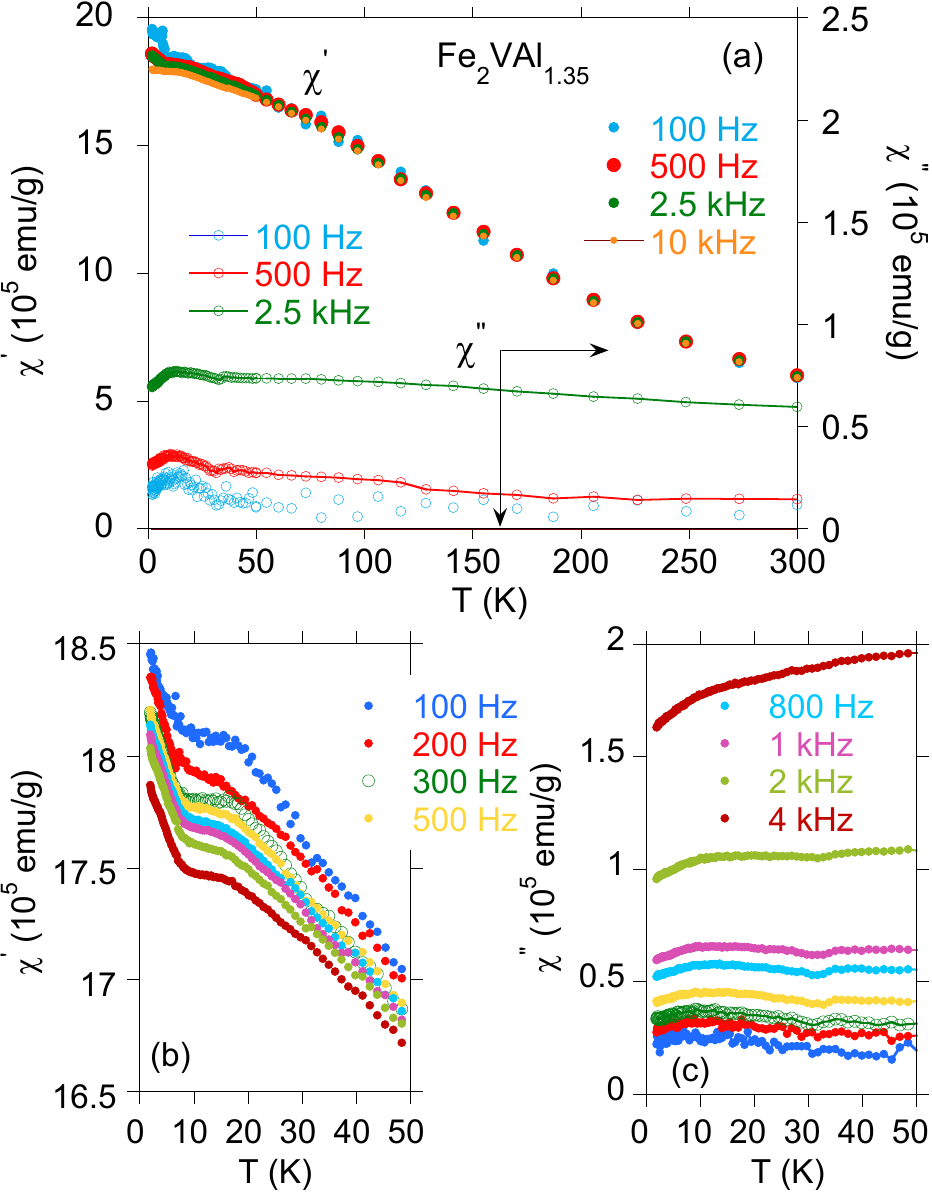}
\caption{\label{fig:Fig_CHI_ac_Fe_A-B-C}
Temperature dependence of the real and imaginary part of the ac magnetic susceptibility $\chi_{ac}$ for Fe$_2$VAl$_{1.35}$ measured for various frequencies of the applied magnetic field; the amplitude of the magnetic field was 2 G.
Panels (b) and (c) show details over a limited temperature range for $T<50$ K. Panel (b) shows the obvious {\it step}-like changes in $\chi'$  at about 20 K, 30 K, and 40 K, which are magnified for clarity in Fig. \ref{fig:CHI_memory}.
}
\end{figure}

The ac magnetic susceptibility was measured at various frequencies in order to confirm the hypothesis of spin/cluster-glass state in Fe$_2$VAl$_{1.35}$.
Shown in Fig.~\ref{fig:Fig_CHI_ac_Fe_A-B-C} are the real ($\chi'$) and imaginary ($\chi''$) components of the magnetic ac susceptibility data. 
$\chi'$ [in panel (b)] and $\chi''$ [in panel (c)] 
exhibit a broad maxima at $\sim 15$~K with amplitudes and positions depending on the frequency $\nu$ of the applied magnetic field. 
The maximum of $\chi'$ can be attributed to a spin-glass-like transition, which is commonly used to determine the spin freezing temperature $T_f$.
The frequency dependence of $T_f$ follows the empirical Vogel-Fulcher relation that 
is described as $\nu = \nu_0 \exp[-E_a/k_B(T_f - T_0)]$, where $\nu_0$, $T_0$, and $E_a$ are fitting parameters \cite{Mydosh1993}. Taking into account the microscopic single spin flipping frequency $\nu_0\sim 10^{13}$ Hz for spin–glass materials we obtained $E_a=290$ K and $T_0=1.7$ K, fitting this expression to the linear change of $T_f$ versus $1/\ln(\nu_0/\nu)$. The parameter $T_0$ does not have a precise physical meaning; it is proposed to be related to the true critical temperature when $T_f>T_0$ is only a dynamic manifestation of the magnetic transition from paramagnetic to SG phase, cf. Ref. \cite{Mydosh1993}.
The frequency shifts of the $\chi'$ maxima yield ratio $\delta T_f=\Delta T_f/T_f\Delta \log_{10}\nu\approx 0.02$, which
is one order of magnitude higher than expected for canonical metallic spin-glass materials ($\sim 10^{-3}$), but fits well with the range that is reported for cluster-glasses \cite{Mydosh1993}.
\begin{figure}[h!]
\includegraphics[width=0.45\textwidth]{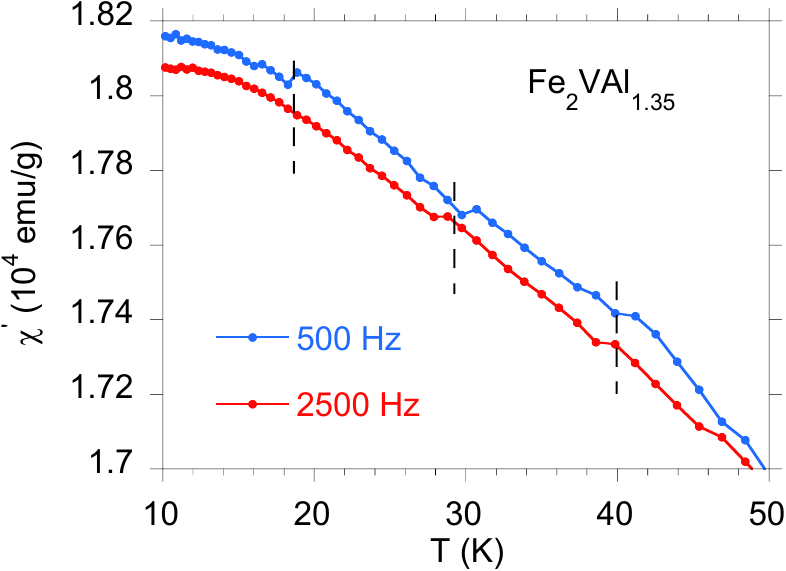}
\caption{\label{fig:CHI_memory}
Magnetic susceptibility $\chi'$ as a function of temperature; drops in $M'$ during stops at 19, 30 and 41 K are magnified for clarity (cf. Fig. \ref{fig:Fig_CHI_ac_Fe_A-B-C}) and will be discussed in Sec. III.B.
}
\end{figure}   
\begin{figure}[h!]
\includegraphics[width=0.45\textwidth]{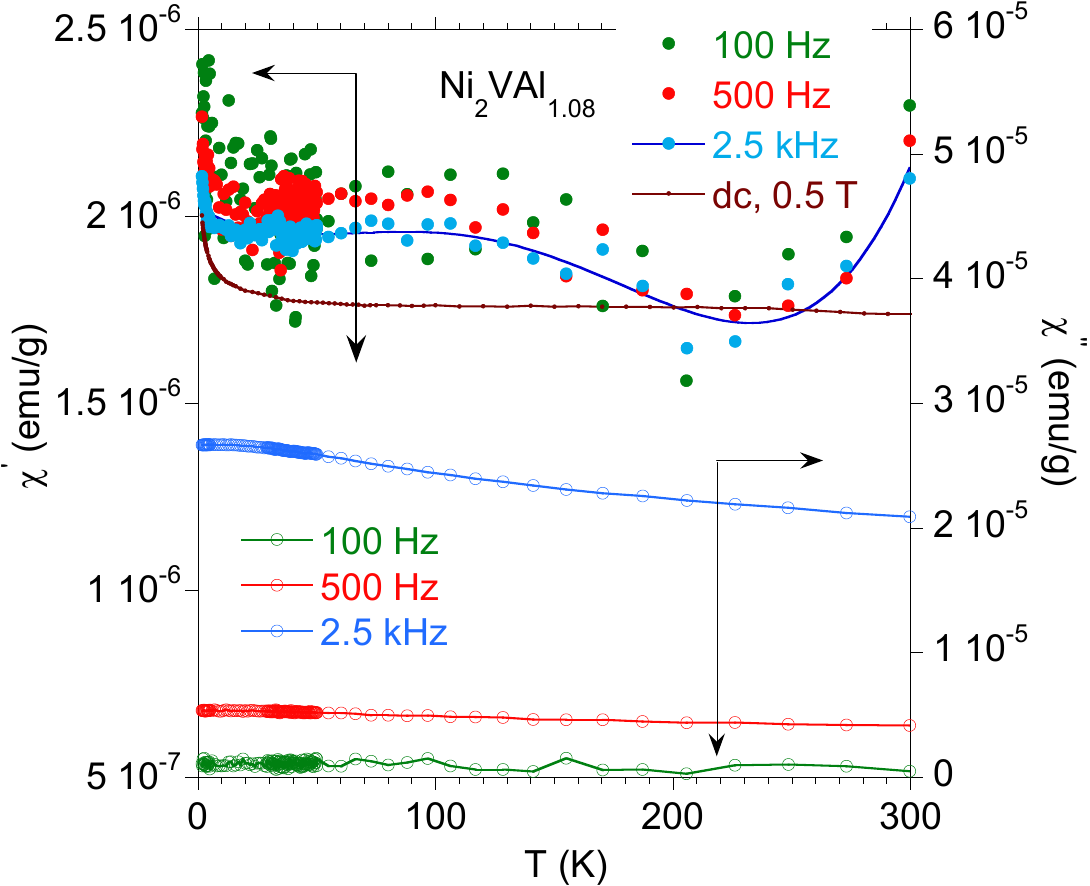}
\caption{\label{fig:CHI1-CHI2_CHI-dc}
Temperature dependence of the real and imaginary part of the ac magnetic susceptibility $\chi_{ac}$ for Ni$_2$VAl$_{1.05}$ measured for various frequencies of the applied magnetic field; the amplitude of the magnetic field was 2 G. Comparison with dc magnetic susceptibility measured for Ni$_2$VAl$_{1.05}$ in the field 5000 G.
}
\end{figure} 
Fig. \ref{fig:CHI1-CHI2_CHI-dc}  shows quite  different behavior  in $\chi_{ac}(T)$  for   Ni$_2$VAl$_{1.08}$, namely, a broad maximum  in $\chi'$  is observed at about 100 K, which is indicative of the spin fluctuations.

\begin{figure}[h!]
\includegraphics[width=0.45\textwidth]{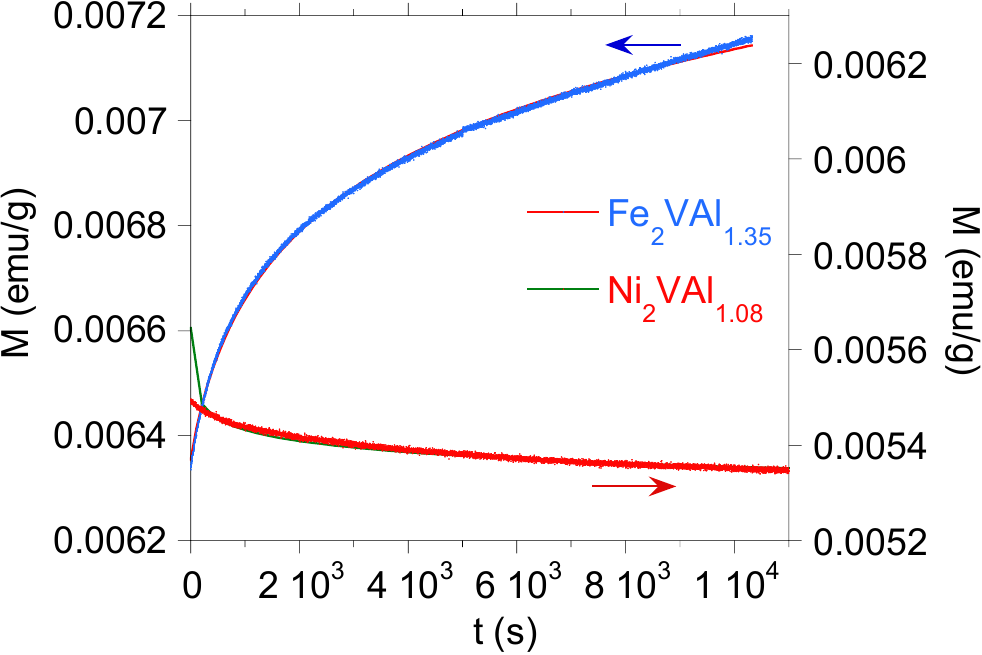}
\caption{\label{fig:Fig_M-t_Fe-Ni}
Time dependent remnant magnetization behavior for Fe$_2$VAl$_{1.35}$ (blue points) and Ni$_2$VAl$_{1.08}$ (red points). The red solid line represents a fit to 
equation $M(t)=M(0)+S\ln(1+t/t_0)$, the green solid line represents the best fit to equation $M(t)= M(0)t^{-\alpha}$.
}
\end{figure}

We have studied the isothermal magnetic relaxation phenomenon in Fe$_2$VAl$_{1.35}$ and  Ni$_2$VAl$_{1.08}$ as a final test of glassy state formation in these alloys. 
The samples were first zero field cooled from 300 K down to 8 K with a constant cooling rate and kept at a target temperature for a waiting time $t_w=300$ s in the field of 5000 G. Then the field was switched off. 
Figure \ref{fig:Fig_M-t_Fe-Ni} displays the time evolution of magnetization $M$ measured in zero-field-cooled (ZFC) mode at
temperature 8 K for an applied field of 0.04 G. 
Various functional forms have been proposed to describe magnetization as a function of observation time \cite{Mydosh1993}.
The time dependence of $M$ shown in Fig. \ref{fig:Fig_M-t_Fe-Ni}  for Fe$_2$VAl$_{1.35}$ is well approximated by the expression for magnetic viscosity
$M(t)=M(0)+S\ln(1+t/t_0)$,
where $M(0)$ is the magnetization at $t=0$, $t_0=375$ s is the reference time, and $S=2.3\times 10^{-4}$ emu/g is the magnetic viscosity. The reference time $t_0$ is typically orders of magnitude larger than the observed microscopic spin flip $\tau_0$. 
The estimated values of $S$ are comparable to the results reported for other glassy systems.
The magnitude of $M(t)$ strongly depends on $t_w$  before switching on the field \cite{Mydosh1993}. However, this behavior is out of the scope of this research.
Alternatively, $M(t)$  can be well approximated by an expression $M(t)=M_0+M_r\exp[-(t/\tau_r)^{1-n}]$, where
magnetization  $M(0)=0.0092$ emu/g could be interpreted as an intrinsic weakly
{\it ferromagnetic} component that appears below $\sim 200$ K in effect of sample disorder (cf. Fig. \ref{fig:CHI_dc_Fe-Ni_inset}), while $M_r=0.0031$ emu/g could be related to a glassy component that mainly contributes to the relaxation effects observed.
Within the disordered scenario, the magnetic clusters of Fe are distributed in the weakly {\it magnetic} background. In this approximation, the time constant $\tau_r=1.8\times 10^5$ s and the parameter $n=0.35$  \cite{comment3} are related to the relaxation rate of the spin-glass-like phase \cite{Freitas2001}.
Figure \ref{fig:Fig_M-t_Fe-Ni} compares a similar  isothermal
remnant magnetization (IRM) as a function of time, measured  for Ni$_2$VAl$_{1.08}$ at 8 K under the same conditions. The observed time dependence of
IRM is weakly $t$-dependent  and
can be fitted by the power-law decay, $M(t)= M(0)t^{-\alpha}$ with the fitting parameter $\alpha = 6.2 \times 10^{-3}$, however, only for $t>200$ s. Below this time limit, the $M(t)$ data do not follow the $M(0)t^{-\alpha}$ behavior.  The $M(t)$ dependence shown  for  Ni$_2$VAl$_{1.08}$  in Fig. \ref{fig:Fig_M-t_Fe-Ni} signals the presence of diluted and disordered magnetic moments of AS Ni defects (will be discussed) which, however, do not form an ordered glassy state.

\begin{figure}[h!]
\includegraphics[width=0.45\textwidth]{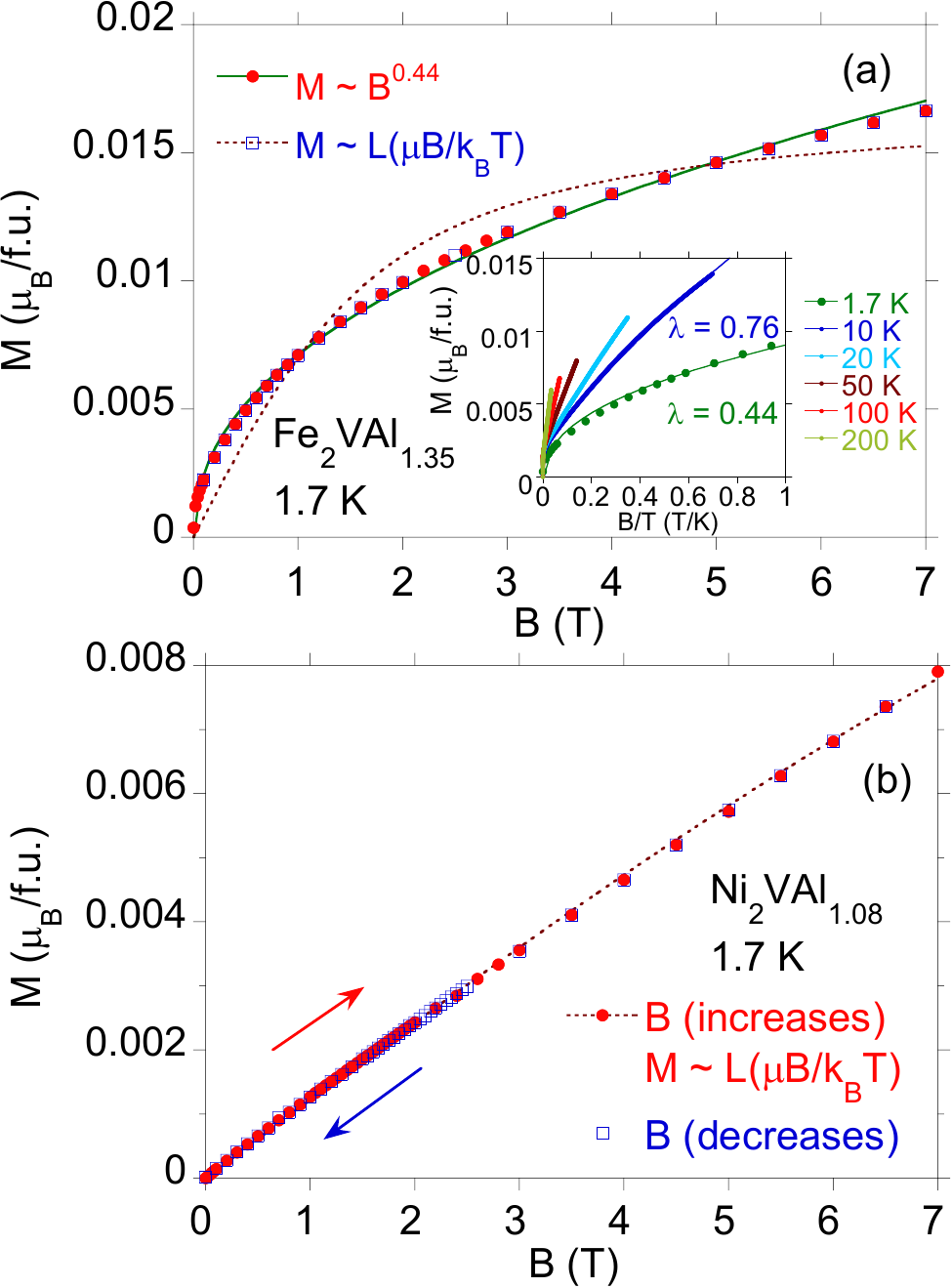}
\caption{\label{fig:M_B_fity_Fe-Ni_A-B_inset}
Isothermal magnetization as a function of applied magnetic field of Fe$_2$VAl$_{1.35}$ (a) and Ni$_2$VAl$_{1.08}$ (b) for 1.7 K. The red dotted line is a fit of the Langevin function $L$ to the
magnetization data. In panel (a) the green line shows the approximation of the expression $M=M(0)+mB^{\lambda}$ to the experimental data. The inset displays $M$ vs $B/T$; in this plot the $M$ isotherms are approximated by the $M\sim B^{\lambda}$ expression for  $T=1.7$ K and 10 K, respectively.   
}
\end{figure}

To further probe the nature of the magnetic ground state in Fe$_2$VAl$_{1.35}$ and Ni$_2$VAl$_{1.08}$ Heusler alloys, the isothermal magnetization $M$ was measured as a function of magnetic field, as shown in Fig. \ref{fig:M_B_fity_Fe-Ni_A-B_inset}.
The $M(B)$  isotherms of Fe$_2$VAl$_{1.35}$  do not exhibit any hysteresis loop, however, are also not characteristic of paramagnets. The $M(B)$ characteristics cannot be
approximated by the Langevin function $L(\xi)=\coth(\xi)-1/\xi$ 
($\xi=\mu B/k_BT$  and $\mu$ is the total magnetic moment), as shown in panel (a). Moreover,  $M$ is not a universal function of $B/T$ as displayed in the inset to Fig. \ref{fig:M_B_fity_Fe-Ni_A-B_inset}(a). This scaling behavior is characteristic of the superparamagnetic state, which is not the case.
Whereas, magnetization as a function of the field up to 7 T follows the predicted Griffiths phase behavior $M\sim B^{\lambda}$ \cite{Castro2000}, as shown in Fig. \ref{fig:M_B_fity_Fe-Ni_A-B_inset}(a). Within the Griffiths phase scenario the $M\sim B^{\lambda}$   
well approximates the experimental data and gives $\lambda=0.76$ for the $M(B)$ data at $T=10$ K. The fitting procedure of $M$ at $T=1.7$~K gives $\lambda\sim 0.45$ smaller than expected, the reason is due to the increase of magnetic correlations for $T\rightarrow T_C^g$ that lead to a glassy state and dominate the Griffiths phase state. 
Figure \ref{fig:M_B_fity_Fe-Ni_A-B_inset}(b) displays the paramagnetic $M$ vs. $B$ behavior for Ni$_2$VAl$_{1.08}$ at 1.7 K, well approximated by the Langevin function $L(\xi)$.

Finally, we present some notes on the magnetic ground state in the disordered Fe alloy. While the parent compound Fe$_2$VAl has a {\it nonmagnetic} ground state, its  disordered analogues determined by the presence of vacancies at various crystallographic sites, off-stoichiometry, and/or doping are very close to ferromagnetic ordering, as has been demonstrated by many studies (cf. Ref. \cite{Tsujii2019} and references therein). The divergence in the $\chi(T)$ data shown in Fig. \ref{fig:CHI_dc_Fe-Ni_inset} suggests this possibility at the limit of $T\rightarrow 0$.

\subsection{Memory effect in Fe$_2$VAl$_{1.35}$}

The non-zero value of $M(0)$ in Fig. \ref{fig:Fig_M-t_Fe-Ni} indicates the coexistence of weakly magnetic and glassy magnetic components in the relaxation process. It can be assumed that the small clusters of Fe are separated in the weakly magnetic phase. The phase separation scenario would be favorable to explain the existence of out-of-equilibrium features shown in $\chi' (T)$ data in Figs. \ref{fig:Fig_CHI_ac_Fe_A-B-C} and \ref{fig:CHI_memory}, as competition between coexisting phases, leading to the appearance of locally metastable states observed in ac susceptibility, as memory effect in the cooling cycle. In our ac magnetization measurements, a field with an amplitude of 2 G was applied during cooling. An analogous behavior was observed for Fe-doped phase separated manganite La$_{0.5}$Ca$_{0.5}$MnO$_3$ \cite{Levy2002} and for Dy$_{0.5}$Sr$_{0.5}$MnO$_3$ \cite{Harikrishnan2010}.

\subsection{Specific heat}

\begin{figure}[h!]
\includegraphics[width=0.45\textwidth]{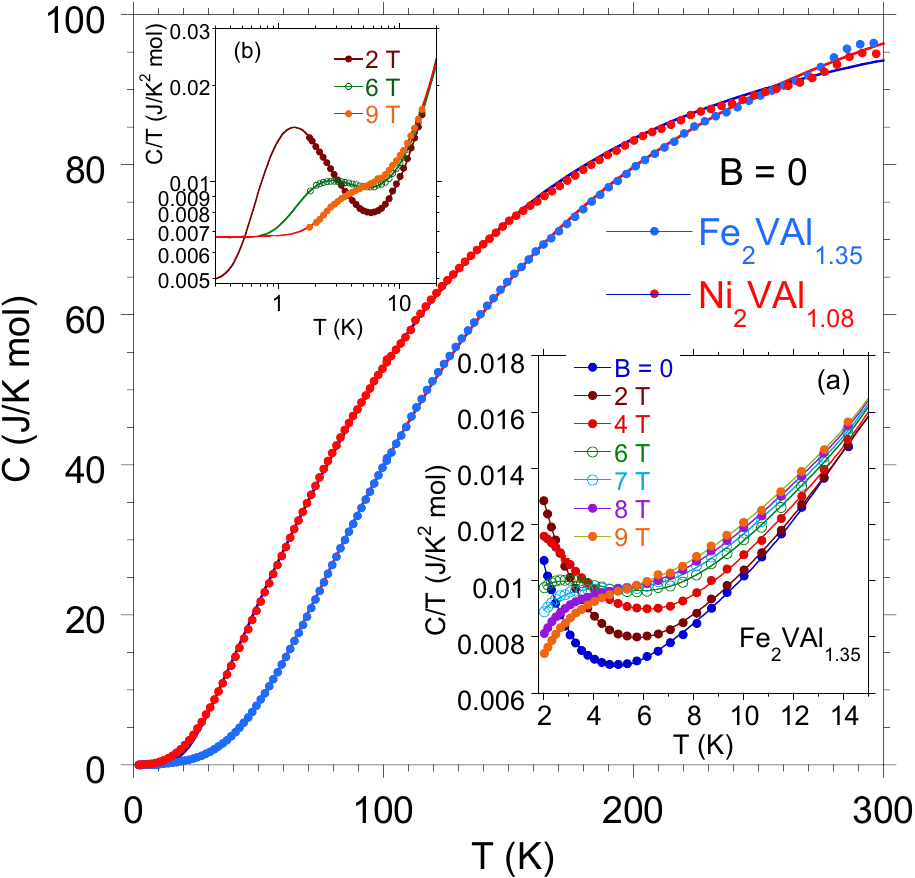}
\caption{\label{fig:Fig_C_Fe-Ni_inset_A-B}
Specific heat for Fe$_2$VAl$_{1.35}$ (blue points) and Ni$_2$VAl$_{1.08}$ (red points) measured
in a wide temperature range and a fit of Eq. (\ref{eq:Debye-Einstein}) to the experimental data. In panel (a), the temperature dependence of $C/T$ in various magnetic fields. The continuous curves are the
fits of a two-level Schottky-like function $C_{\rm Sch}$, lattice $\beta T^3$, and electronic $\gamma T$ contributions to the $C/T$
data (cf. Table \ref{tab:Table_C_Sch}). The inset (b) shows in detail the approximation of the two-level Schottky function to the experimental data $C/T$ for Fe in the field of 2, 6, and 9 T.
}
\end{figure}

Fig.~\ref{fig:Fig_C_Fe-Ni_inset_A-B} compares the temperature dependence of the specific heat for Fe$_2$VAl$_{1.35}$ and of Ni$_2$VAl$_{1.08}$, measured in a zero magnetic field.  For both alloys, the value of $C$ per one atom reaches the value of $3R$ in accordance with the Dulong-Petit law ($R$ is the gas constant). The $C(T)$ data are well approximated by the Debye-Einstein (DE) model \cite{Tari2003}:
\begin{eqnarray}
C (T) &=& \gamma_0 T + 9 N R(1-d) \left( \frac{T}{\Theta_{\rm D}} \right)^3 \int_0^{\Theta_{\rm D}/T} \frac{x^4 \; e^x}{(e^x - 1)^2} dx  \nonumber\\ 
&&+\: 3nRd\left(\frac{\theta_E}{T}\right)^2\frac{e^{\theta_E/T}}{(e^{\theta_E/T - 1})^2}, 
\label{eq:Debye-Einstein}
\end{eqnarray}
where the first term is the electron specific heat $C_{el} (T)=\gamma_0 T$, and the two others account for the lattice contributions ($\theta_D$ and $\theta_E$ are the Debye and Einstein temperatures, respectively, $n$ is the number of atoms per formula unit, and $d$ stands for the number of optical phonon modes). 

The solid lines show temperature variation of the calculated $C(T)$ with the fitting parameters $\gamma_0 =33$ mJ/molK$^2$,  $\theta_D =638$ K, $\theta_E =297$ K, and $d=0.51$ for Fe$_2$VAl$_{1.35}$, and respective  set of the fitting parameters for Ni$_2$VAl$_{1.08}$ ($\gamma_0 =11$ mJ/molK$^2$, $\theta_D=492$ K, $\theta_E=148$ K, and $d=0.69$).  
Equation (\ref{eq:Debye-Einstein}) does not take into account the magnetic contributions from the spin glass state, therefore, at the temperatures $T<20$ K the fitting is not satisfactory for Fe$_2$VAl$_{1.35}$, which is the reason for the overestimated value of $\gamma_0$.

The $\gamma_0$ and $\beta_0$ derived from the linear sections (for $T>6$ K) of $C/T$ vs $T^2$ dependence are 5.5 mJ/K$^2$ mol and $4.6\times 10^{-5}$ J/K$^4$ mol, respectively, indicating the pseudogap in the bands of Fe$_2$VAl$_{1.35}$ at the Fermi level. 
The $\gamma_0$ and $\beta_0$  determined similarly for Ni$_2$VAl$_{1.08}$ in the temperature range $T<15$ K are  respectively, 13.5 mJ/K$^2$ mol and 2.6 J/K$^4$ mol. Moreover, these fitting parameters obtained from the approximation of Eq. (\ref{eq:Debye-Einstein}) to $C(T)$ as well those from linear dependence of $C/T$ vs. $T^2$ are similar.
For $N=4$ atoms in formula unit,  $\beta =N(12/5)\pi^4R\theta_D^{-3}$ gives the Debye temperature, $\theta_D \sim 567$ K for Fe$_2$VAl$_{1.35}$ and $\theta_D \sim 304$ K for Ni$_2$VAl$_{1.08}$, respectively. In both cases, the determined temperatures $\theta_D$ are close to those obtained by fitting the DE function [Eq. (\ref{eq:Debye-Einstein})] to the $C(T)$ data. 

Inset (b) displays the low-temperature specific heat $C$ divided by temperature, $C(T)/T$, at various magnetic fields for Fe$_2$VAl$_{1.35}$.
The upturn in $C/T$ at $B\neq 0$ can be interpreted as a result of Schottky-like anomalies due to magnetic defects \cite{Lue99,Lue05}.
Assuming that the specific heat is a sum, $C_{\rm magn} + \gamma T + \beta T^3$, the $C/T$ data at various magnetic fields are well approximated to expression $C/T =C_{\rm Sch}/T + \gamma  +\beta T^2$, where $C_{\rm Sch}$ is a two-level Schottky function 
\begin{equation}
C_{\rm Sch}=N_{Fe}k_B\left(\dfrac{\epsilon}{T}\right)^2\dfrac{e^{\epsilon/T}}{(1+e^{\epsilon/T})^2},
\label{eq:Schottky}
\end{equation}
 with $\epsilon$, $\gamma$, and $\beta$ field dependent (cf. Table \ref{tab:Table_C_Sch}). 
In the inset (b) to Fig.~\ref{fig:Fig_C_Fe-Ni_inset_A-B} the solid lines are the best fits of $C_{\rm Sch}/T + \gamma  +\beta T^2$ to the $C/T$ experimental data of Fe$_2$VAl$_{1.35}$. A simple fit of the Schottky function to $C(T,B=0)$ data gives
$\sim 10\times 10^{19}$ cm$^{-3}$ of Schottky centers $N_{Fe}$. 
Assuming that Fe$_{AS}$ defects dominate in the sample, and taking $g=1.93$ and $S=3/2$ for the antisite defects, the saturation magnetization of Fe$_2$VAl$_{1.35}$  (cf. Fig. \ref{fig:M_B_fity_Fe-Ni_A-B_inset}$a$) expressed  by $M_S=N_{Fe}g\mu_B S$ gives comparable {\it magnetic impurity} concentration of $\sim 13\times 10^{19}$ cm$^{-3}$ which, when calculated per unit cell, gives about 0.02 atom per cell in the antisite position.
One also notes that the $\epsilon$ vs. $B$ change is well approximated by expression $\epsilon(B)=\epsilon(0) + bB^n$ with exponent $n=1.6$, moreover, in the constant field mode, each value of $\epsilon/k_B$ is obtained twice larger than the value of 
$T\equiv T_f$ at  which  $C_{B={\rm const}}(T)/T$  has the maximum. 
\begin{table*}[h!]
\caption{Specific heat $C(T)$ parametrization of Fe$_2$VAl$_{1.35}$  within temperatures $T<15$ K. $C(T)$ is fitted to the two-level Schottky function supplemented with electron ($\gamma T$) and phonon ($\beta T^3$) contributions. Comparison with electronic specific heat coefficients $\gamma_0$ and coefficients $\beta_0$ experimentally
obtained from a least-squares fit of expression $C(T)/T=\gamma_0 +\beta_0 T^2$ to the experimental data in the temperature region between $\sim 5$ K and $15$ K (cf. Fig. \ref{fig:Fig_gamma-beta_Fe_Ni_sum}).
}
\label{tab:Table_C_Sch}
\begin{tabular}{cccccc}
\hline\hline
& &  $C=N_{Fe}k_B(\epsilon/T)^2e^{\epsilon/T}/(1+e^{\epsilon/T})^2 +\gamma T +\beta T^3$   &  & $C=\gamma_0 T +\beta_0 T^3$ & \\

$B$ (T)  & $\epsilon$ (K) & $\gamma$ (mJ/K$^2$ mol)&  $ \beta\times 10^{5}$ (J/K$^4$ mol) & $\gamma_0$ (mJ/K$^2$ mol)&  $ \beta_0\times 10^{5}$ (J/K$^4$ mol) \\
\hline
0 &   2    &  4.7  &  5.0  & 5.5 & 4.6 \\        
2 &  3.2   &  4.9  &  4.8  & 6.0 & 4.4  \\ 
4 &  4.7   &  5.7  &  4.5  & 6.9 & 4.0 \\ 
6 &  6.2   &  6.7  &  4.1  & 7.7 & 3.8  \\    
7 &  7.5   &  7.0  &  4.8  & 8.0 & 3.6 \\    
8 &  9.0   &  7.1  &  4.0  & 8.3 & 3.6\\  
9 &  10.8  &  6.8  &  4.2  & 8.5 & 3.5 \\        
\hline\hline
\end{tabular} 
\end{table*}
\begin{figure}[h!]
\includegraphics[width=0.45\textwidth]{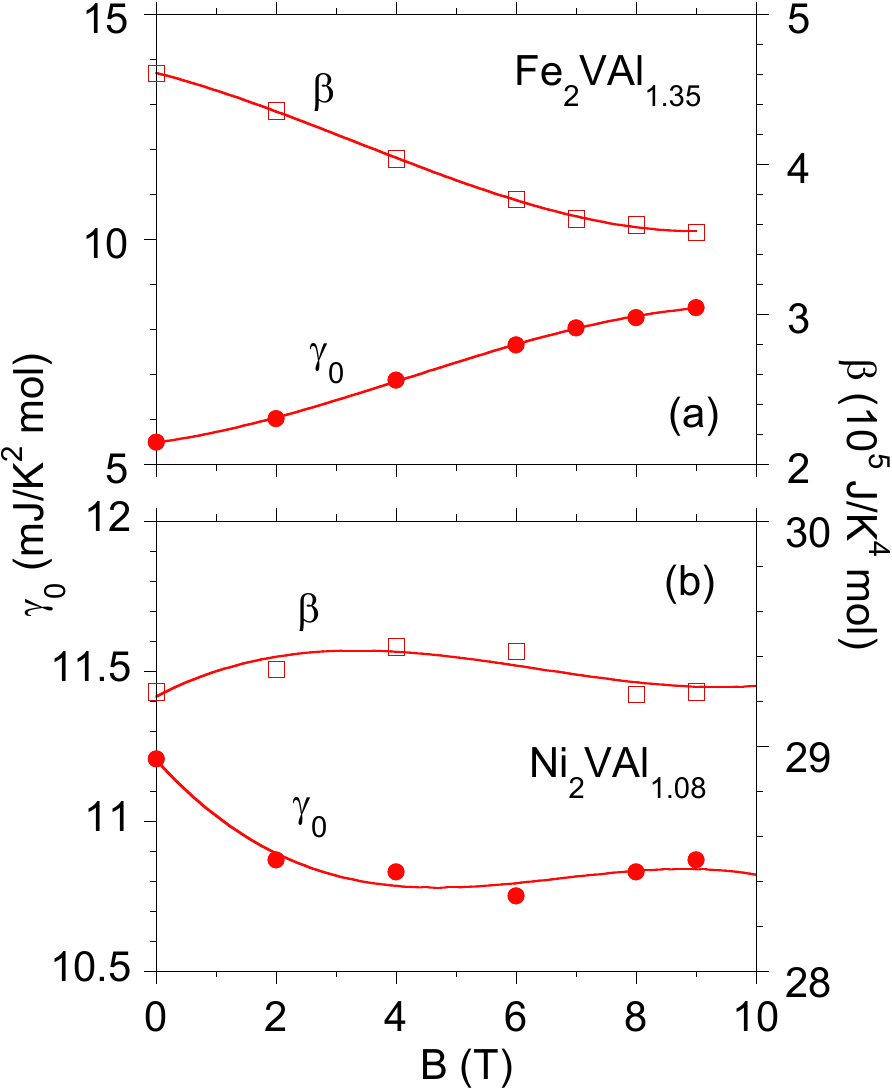}
\caption{\label{fig:Fig_gamma-beta_Fe_Ni_sum}
Electronic specific heat coefficient $\gamma_0$ and coefficient $\beta$  vs applied magnetic field $B$. $\gamma_0$ and $\beta$  are obtained from the linear dependence of $C(T)/T=\gamma_0+\beta T^3$ vs. $T^2$ for Fe$_2$VAl$_{1.35}$  and for Ni$_2$VAl$_{1.08}$ in the temperature region between $\sim 5$ K and $15$ K.
}
\end{figure}

Figure \ref{fig:Fig_gamma-beta_Fe_Ni_sum} shows the electronic specific-heat constant $\gamma_0$ and the coefficient of the $T^3$ term $\beta$ for Fe and Ni samples as a function of the magnetic field.
It is worth noting that the field dependencies either of $\gamma_0$ and $\beta$ are for Ni$_2$VAl$_{1.08}$ typical for systems with spin fluctuations. 
Namely, the field-induced behavior of $\gamma_0$ and $\beta$  shown in panel (b) as well as $\chi(T)$  shown in Fig. \ref{fig:CHI1-CHI2_CHI-dc} are similar to those, observed for canonical spin fluctuator  CeSn$_3$, which was classified by Ikeda et al.  \cite{Ikeda1991} as a type 3 spin fluctuator. 

Fe$_2$VAl$_{1.35}$  exhibits significantly different corresponding characteristics, as shown in panel (a).
An increase in $\gamma_0$ shown in Fig. \ref{fig:Fig_gamma-beta_Fe_Ni_sum} we attribute to a strong reducing of the pseudogap at the Fermi level by increasing the field, which consequently gives an increase of $\gamma$.
Our previous research of similar Heusler alloys (Fe$_2$TiSn \cite{Slebarski2006}, Fe$_2$VSn \cite{Slebarski2004}) explicitly documented that
the physical properties of these semiconductors, in particular resistivity, susceptibility, and specific heat are dominated by crystallographic disorder. The AS atomic disorder
can generate the narrow $d$-electronic  band located at the Fermi level, which is responsible for the unusual temperature dependencies (these materials can also be discussed as false Kondo insulators). 

\begin{figure}[h!]
\includegraphics[width=0.45\textwidth]{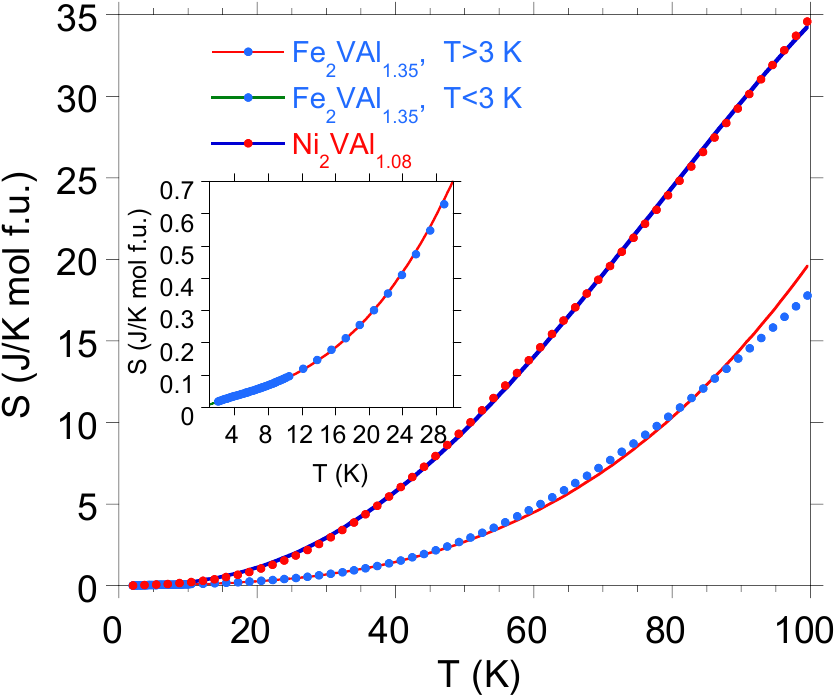}
\caption{\label{fig:S_Fe_Ni_per_FU}
Ni$_2$VAl$_{1.08}$, the fit of Eq. (\ref{eq:entropy}) to entropy $S$ (blue line).  For Fe$_2$VAl$_{1.35}$ entropy is well approximated by expression $S(T) \sim \gamma_0T+\beta T^3+sT^n$ with exponent $n=3/2$ for $T<3$ K (green line) and $n=-2$ for $T>3$ K (red line). The inset displays details.}
\end{figure}

\begin{figure}[h!]
\includegraphics[width=0.45\textwidth]{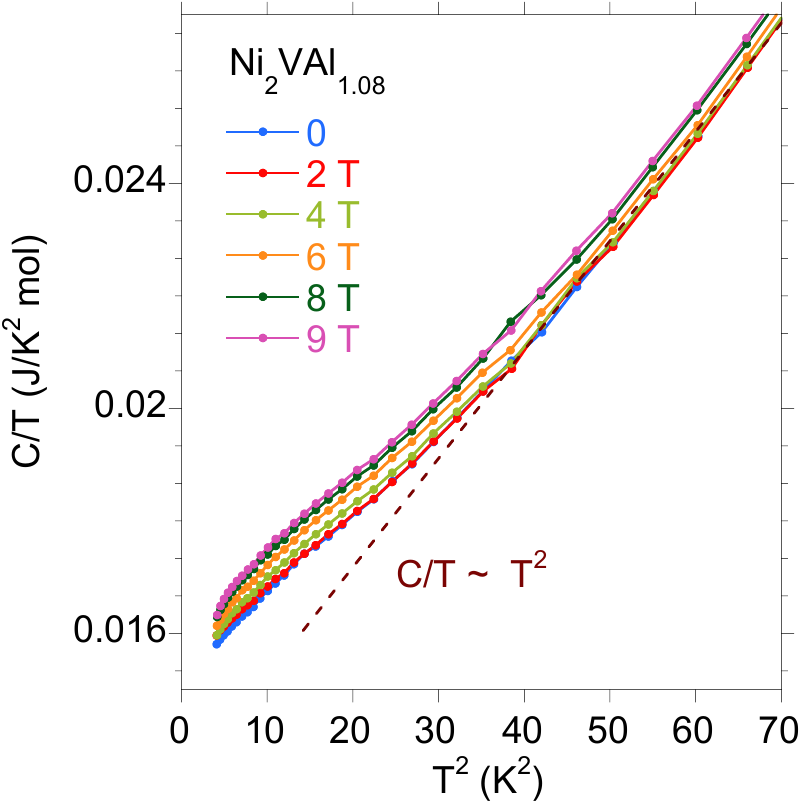}
\caption{\label{fig:Fig_C_T_0-9T_Ni}
Heat capacity $C/T$ vs $T^2$ for Ni$_2$VAl$_{1.08}$  in magnetic fields. The linear $C/T$ vs $T^2$ behavior is  obeyed for temperatures $6<T<20$ K.
}
\end{figure}

Figure \ref{fig:S_Fe_Ni_per_FU} shows entropy $S$ for Fe$_2$VAl$_{1.35}$ and Ni$_2$VAl$_{1.08}$ in the temperature region $T<100$ K.
Assuming  that only conduction electrons, phonons and spin fluctuations contribute to $S$, then the entropy of Ni$_2$VAl$_{1.08}$ can be well fitted by expression  \cite{Spalek2000}:
\begin{eqnarray}
S=\int_{0}^{T}\frac{dC}{T}&=&
\int_{0}^{T}\frac{\gamma_0T+\beta T^3+\delta T^3 \ln(T/
T^{\star})}{T}dT \nonumber \\ 
 &=& \gamma_0T+\frac{\beta}{3} T^3 + \frac{\delta}{3} T^3\ln\frac{T}{T^{\star}}-\frac{\delta}{9}T^3, \hspace*{6mm}
\label{eq:entropy}
\end{eqnarray}
where $T^{\star}=120$ K, $\gamma_0=13$ mJ/K$^2$ mol, $\beta=1.5\times 10^{-4}$ J/K$^2$ mol, and  $\delta=-1.7\times 10^{-4}$ J/K$^4$ mol (cf. Fig. \ref{fig:S_Fe_Ni_per_FU}). The fit is very good for $T>30$ K, but for lower temperatures this approximation deviates from the experimental data, even within 10\% around 15--20 K. A possible reason for this divergence may be the Kondo effect due to scattering of conduction electrons on magnetic Ni impurities, which can give an additional contribution to $S$ (in Sec. IV we will document in {\it ab initio} calculations that Ni at AS positions has a localized  magnetic moment) . 
Figure \ref{fig:Fig_C_T_0-9T_Ni}
presents low-temperature $C/T$ vs. $T^2$ data at
various fields  for Ni$_2$VAl$_{1.08}$, with an obvious and field-dependent upturn in $C/T$ for $T^2<40$ K. 
This behavior is not typical of spin fluctuators for which the quenching of the magnetic contribution to the heat capacity by the magnetic field is usually detected. The possible explanation for the low-temperature heat capacity enhancement shown in Fig. \ref{fig:Fig_C_T_0-9T_Ni} can be the formation of a Kondo resonance. 
Indeed, the heat capacity $C/T$ measured in varying magnetic fields shows typical behavior of diluted Kondo systems \cite{Schotte1975,Degrandes1982}, and well correlates with the low-$T$  resistivity data (see Sec. III.C).

Expression (\ref{eq:entropy}), however, does not fit well the $S$ data of Fe$_2$VAl$_{1.35}$. In this case, the entropy shows a kink at $T\sim 3$ K due to freezing of the glassy phase (see the inset to Fig. \ref{fig:S_Fe_Ni_per_FU}), and can be well approximated by expression $S(T) \sim \gamma_0T+\beta T^3+sT^n$ with exponent $n=3/2$ for $T<3$ K and $n=-2$ for $T>3$ K \cite{Sereni1994}, respectively,  as shown in Fig. \ref{fig:S_Fe_Ni_per_FU} (for both cases, $\gamma_0=5.5$ mJ/K$^4$ mol, $\beta=1.9\times 10^{-4}$ mJ/$K^4$ mol).

\subsection{Electron transport properties}

Previous reports have indicated that even a small deviation in the stoichiometry of Fe$_2$VAl has a significant impact on its electron transport properties \cite{Nishino2001}. Similarly, the thermal heat treatment of this alloy has a decisive impact on its electric and thermal transport \cite{Garmroudi2022}.
So far, electron transport investigations were focused on the Fe$_2$VAl alloys with a deficiency of both Fe, V and Al, mainly in terms of enhancing the thermoelectric properties. 
The aim of the current research was to demonstrate to what extent an excess of Al can change the thermoelectric properties of Fe$_2$VAl.
However, our research did not confirm the expectations of significant strengthening of the thermoelectric properties of this alloy. The results obtained for Fe$_2$VAl$_{1+\delta}$ were compared with electron transport measurements for Ni$_2$VAl.
\begin{figure}[h!]
\includegraphics[width=0.45\textwidth]{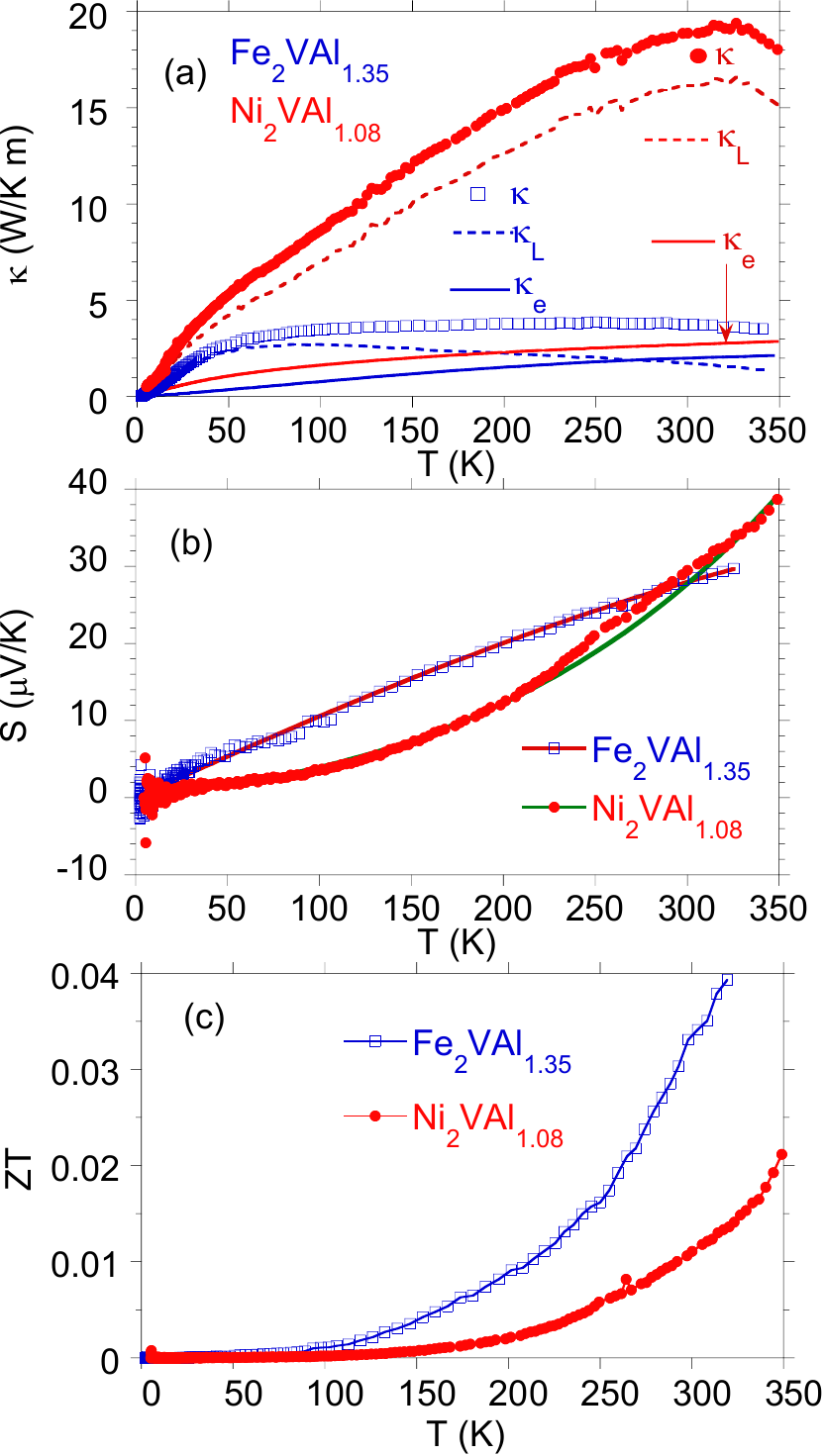}
\caption{\label{fig:K_S_ZT_sum}
Temperature variation of thermal conductivity $\kappa$ (a), Seebeck coefficient $S$ (b), and figure of merit $ZT$ (c) of Fe$_2$VAl$_{1.35}$ (blue open squares) and for Ni$_2$VAl$_{1.08}$ (red points). Panel (a) also shows lattice (dotted line) and electronic (solid line) thermal conductivitiy for Fe$_2$VAl$_{1.35}$ (blue) and Ni$_2$VAl$_{1.08}$ (red), respectively. Panel (b) shows the fit of the SF model (Eq. \ref{eq:S}) to the $S(T)$ data for Fe$_2$VAl$_{1.35}$ (red solid line) and Ni$_2$VAl$_{1.08}$ (blue solid line). The experimental  data of Ni$_2$VAl$_{1.08}$ deviate from the fit around $\sim 270$ K (broad maximum).
}
\end{figure}  
Figure \ref{fig:K_S_ZT_sum} compares the thermal conductivity $\kappa$, Siebeck coefficient $S$, and figure of merit $ZT$ of Fe$_2$VAl$_{1.35}$ and Ni$_2$VAl$_{1.08}$. Panel (a) shows $\kappa$, which is a sum of electronic ($\kappa_e$) and lattice ($\kappa_L$) contributions, measured between 2 and 350 K. 
In general, metals with higher Debye temperatures tend to have higher thermal conductivities. Therefore, one could expect a higher thermal conductivity for Fe$_2$VAl$_{1.35}$  than for its Ni$_2$VAl$_{1.08}$ analogue.
Fe$_2$VAl$_{1.35}$ seems to be, however, an exception to this rule due to the presence of the pseudogap formed at the Fermi level due to interband hybridization, sample off-stoichiometry,  and larger concentration of its antisite defects, all of which contribute to lowering the thermal conductivity of Fe$_2$VAl$_{1.35}$ in respect to $\kappa$ of  metallic Ni$_2$VAl sample.

The electronic thermal conductivity  was evaluated using the Wiedemann-Franz law: $\kappa_e \rho/T = L_0$,  where $\rho$ is the measured dc electrical resistivity and $L_0=2.45\times 10^{-8}$ W$\Omega$K$^{-2}$ is the Lorenz number, while $\kappa_L$ was obtained by subtracting $\kappa_e$ from the measured $\kappa$. The temperature dependencies of $\kappa_L$ and $\kappa_e$ shown in Fig. \ref{fig:K_S_ZT_sum}(a) are typical of disordered crystalline materials where phonon scattering by defects and grain boundaries dominates. 

As shown in Fig. \ref{fig:K_S_ZT_sum}(b), Seebeck coefficient $S$ obtained for  Ni$_2$VAl$_{1.08}$ is
approximated by Eq. \ref{eq:S}, which expresses the temperature dependence of $S_{SF}(T)$  for nearly ferromagnetic, spin fluctuating metals \cite{Okabe2010} such as YCo$_2$ \cite{Gratz2001}, under the condition that their susceptibility $\chi(T)$ shows a broad maximum (cf. Fig. \ref{fig:CHI1-CHI2_CHI-dc}),  
\begin{eqnarray}
S_{SF}=\tilde\alpha T+\tilde \beta T\left(\frac{T}{\tilde T_0}\right)\log\frac{\tilde\delta + (T/\tilde T_0)^2}{(T/\tilde T_0)^2}.
 \hspace*{6mm}
\label{eq:S}
\end{eqnarray}
Within this modeling, $d$ electrons are responsible for the spin fluctuation, while transport properties are due to conduction electrons, which are dragged by spin fluctuations (SF), $\tilde\alpha$, $\tilde\beta$, $\tilde\delta$, and $\tilde T_0$ are fitting parameters. The fitting procedure gives $\tilde\alpha=0.029$ $\mu$V K$^{-2}$, $\tilde\beta=0.053$ $\mu$V K$^{-2}$, $\tilde T_0=220$ K, and $\tilde\delta=4.3$ for Ni$_2$VAl$_{1.08}$. The experimental data for $S(T)$ of Fe$_2$VAl$_{1.35}$ can also be approximated by expression (4) with the fitting parameters $\tilde\alpha=0.108$ $\mu$V K$^{-2}$, $\tilde\beta=-0.021$ $\mu$V K$^{-2}$, $\tilde T_0=300$ K,  and $\tilde\delta=4$. Nonetheless, spin fluctuations do not significantly enhance the value of figure of merit, $ZT$, which for the both alloys is about $10^{-2}$ at 350 K, as shown in Fig. \ref{fig:K_S_ZT_sum}(c).

\begin{figure}[hbt!]
\includegraphics[width=0.45\textwidth]{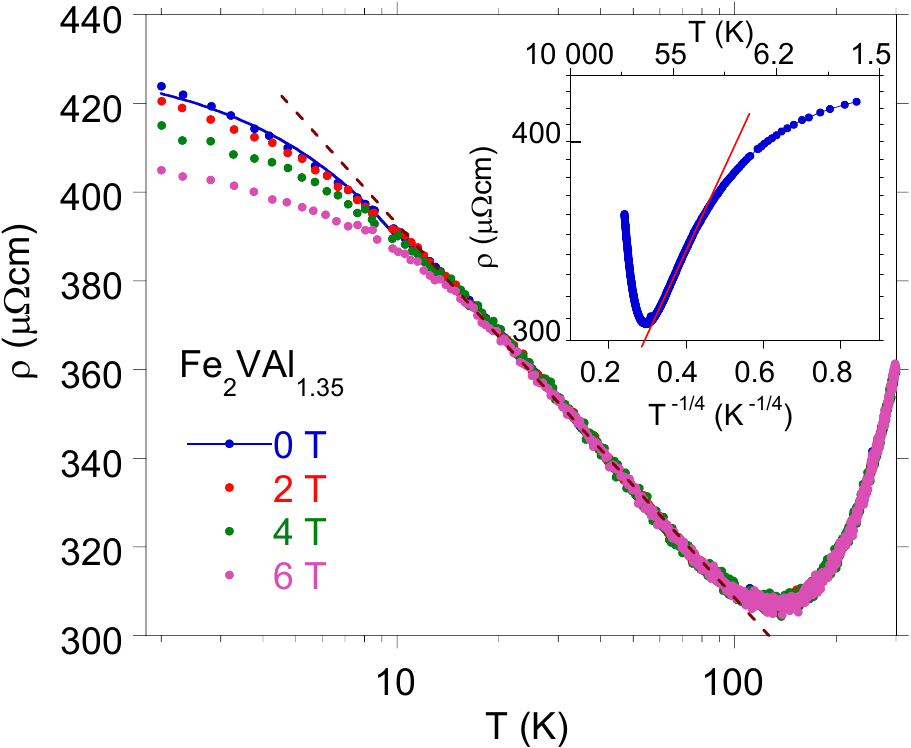}
\caption{\label{fig:Fig_Fe_RvsT_H_inset}
Resistivity vs. temperature (ln$T$ scale) of Fe$_2$VAl$_{1.35}$ in magnetic fields up to 6 T. The dashed line indicates the ln$T$ behavior  in  $\rho (T)$. The expression $\rho(T)=\rho_0+bT^5+c\ln(\mu/T)+\rho_{vrh}\exp(T_0/T)^{1/4}+r_{SG}T^{3/2}$ fits the $\rho(T,B=0)$ data between 2 and 100 K (blue line), where the respective  parts  are  the phonon, Kondo, Mott-VRH, and spin-glass contributions to the resistivity. The inset shows $\ln\rho$ vs. $T^{-1/4}$, with the linear change of $\rho$ marked by the red line.
}
\end{figure}

In Fig. \ref{fig:Fig_Fe_RvsT_H_inset} we present the resistivity $\rho$ of  Fe$_2$VAl$_{1.35}$ at different magnetic fields. 
The $\rho(T)$ data deviates from those, usually reported for stoichiometric Fe$_2$VAl compound \cite{Nishino1997,Slebarski2006a} and exhibit a semiconducting-like behavior, similar to that,  reported for  Fe$_2$TiSn \cite{Slebarski2000,Chaudhuri2019} and its alloys \cite{Slebarski2004}.
Between $\sim100$ and 10 K $\rho\sim -lnT$, which is usually characteristic of Kondo behavior, whereas a significant deviation from linearity is observed below 10 K. However, we note that within this low-$T$ range there is observed a complex glassy-like phase, which complicates the interpretation of the $\rho$-data.
Moreover,  for $30<T<100$ K conductivity follows $\sigma=\sigma _{vrh}\exp[-(T_0/T)^{1/4}]$ (see Fig. \ref{fig:Fig_Fe_RvsT_H_inset}, inset), which is typical for Mott variable-range hopping (VRH) behavior in three dimensions \cite{Mott1966,Mott1974}. Here $T_0$ characterizes the pseudogap at the Fermi level in the case of solids, where a conduction and valence band overlap giving a finite density of states $DOS(\epsilon_F)$ (see Sec. IV). With increasing overlap of the bands, mostly the $d$-electron states become delocalized, which can lead to a metal-insulator transition of Anderson type. In the limit of weak localization, conduction by hopping (VRH) could be possible, this is a case of the Fe sample. 
$T_0$ inversely depends on the localization length $\xi_L=[\frac{1}{18}DOS(\epsilon_F)k_BT_0]^{-1/3}$, which diverges at the insulator-metal transition \cite{Delahaye1998}.
The best approximation of $\rho= \rho_{vrh}\exp[(T_0/T)^{1/4}]$ to the experimental data shown in Fig. \ref{fig:Fig_Fe_RvsT_H_inset} gives $ \rho_{vrh}=208.7$ $\mu\Omega$cm and $T_0=2.3$ K. Then, the localization length $\xi_L\approx 260$ \AA~for $DOS(\epsilon_F)=1.078$ states/eV (cf. Table \ref{tab:TableDFT}) is quite large, indicating direct proximity to the insulator-metal transition.  Due to the above, we suggest that the hopping between Anderson-localized states is possible. Very recently, a similar conclusion was presented for stoichiometric  Fe$_2$VAl  \cite{Garmroudi2022}.
\begin{figure}[hbt!]
\includegraphics[width=0.45\textwidth]{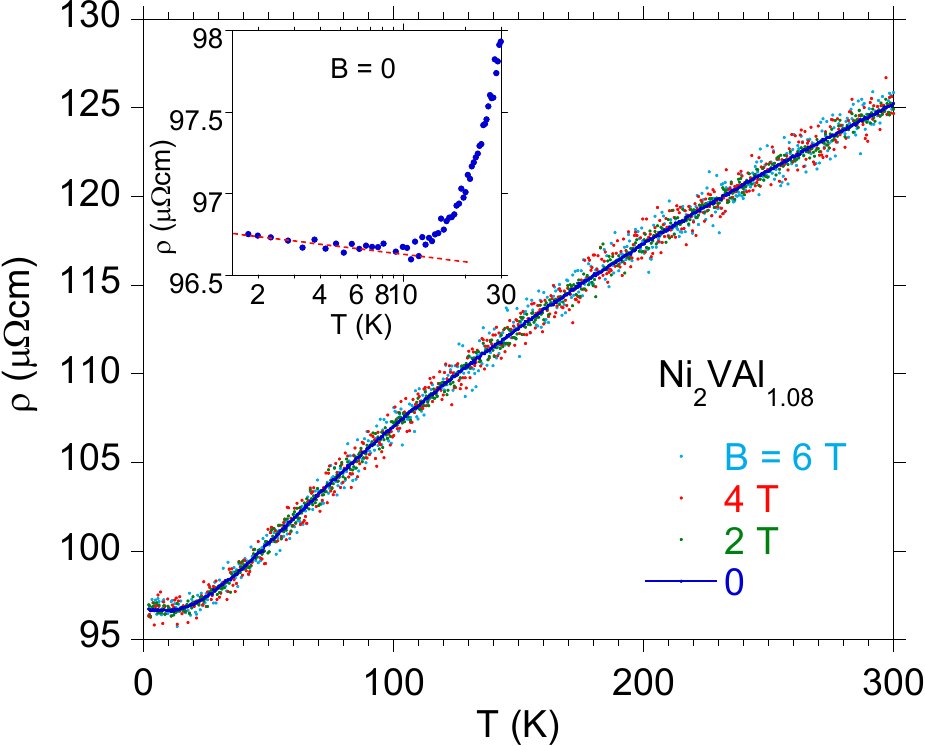}
\caption{\label{fig:Fig_Ni_R-inset}
Resistivity of Ni$_2$VAl$_{1.08}$ vs temperature in applied magnetic fields. In applied magnetic fields $\rho$ vs $T$ is measured with chaotic distribution of the experimental points with respect to the curvature of $\rho(T)$ for zero field. This chaotic spread of experimental points increases with increasing $B$. The inset shows a Kondo impurity behavior $\rho\sim -lnT$.
}
\end{figure}  

The electrical transport of  Fe$_2$VAl$_{1.35}$ reflects its complex interband and magnetic interactions; therefore, the resistivity of this alloy is discussed in relation to the paramagnetic Ni$_2$VAl$_{1.08}$. Figure \ref{fig:Fig_Ni_R-inset} displays the resistivity of Ni$_2$VAl$_{1.08}$ as a function of temperature and in various magnetic fields. The $\rho(T)$  shown in the figure is almost not field-dependent. 
What is interesting, at the lowest temperatures $\rho\sim -\ln T$, suggesting the scattering mechanism of conduction electrons due to the presence of magnetic Ni impurities (Kondo impurity effect).
\begin{figure}[hbt!]
\includegraphics[width=0.45\textwidth]{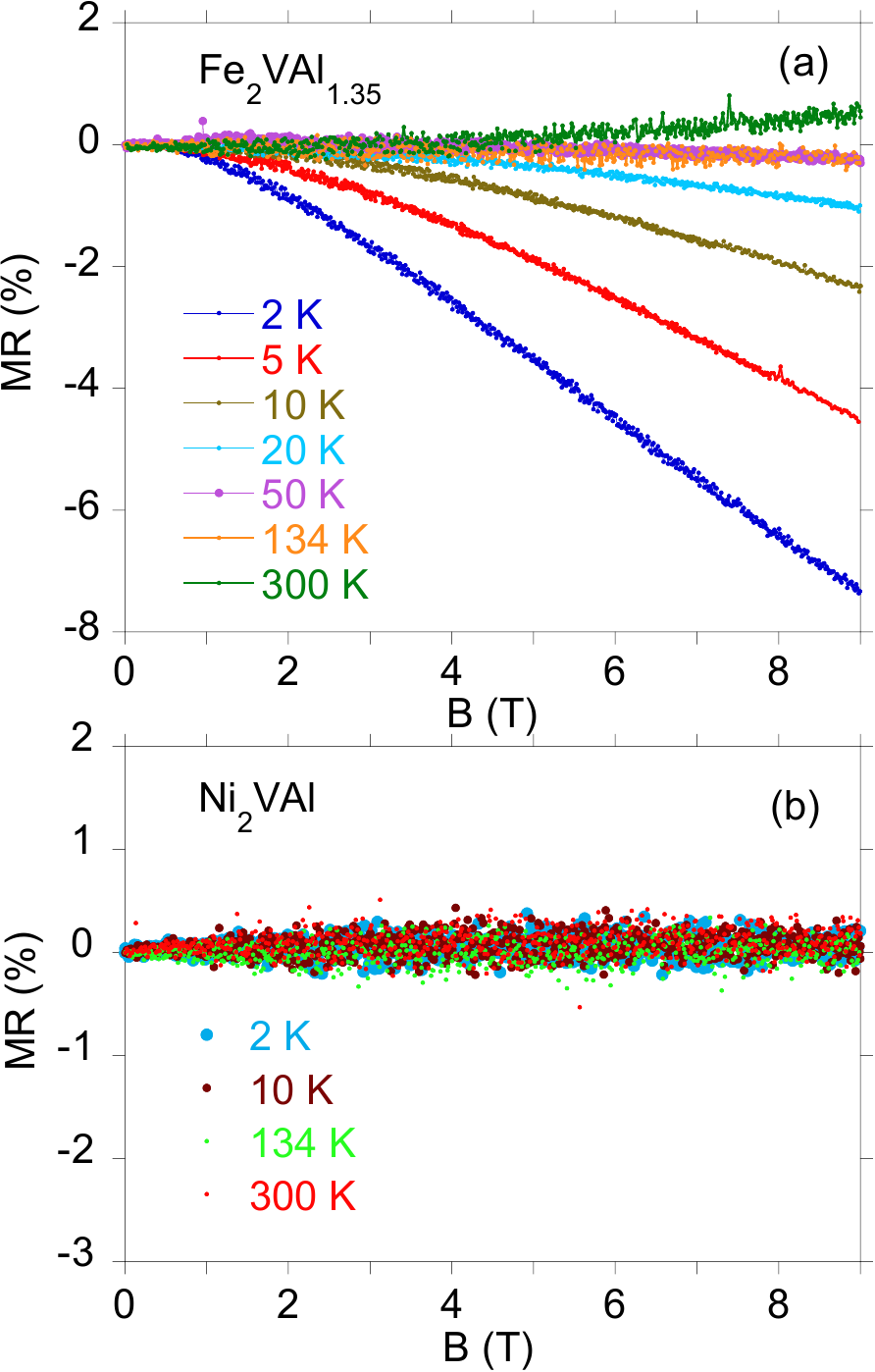}
\caption{\label{fig:Fig_MR_Fe-Ni}
(a) Magnetoresistance $MR = [\rho(B)- \rho(B = 0)]/\rho(B = 0)\times 100$\% ($I=5$ mA), of Fe$_2$VAl$_{1.35}$ (a) and Ni$_2$VAl$_{1.08}$ (b) as a function of the magnetic field $B$, measured at several temperatures. 
}
\end{figure} 
\begin{figure}[h!]
\includegraphics[width=0.45\textwidth]{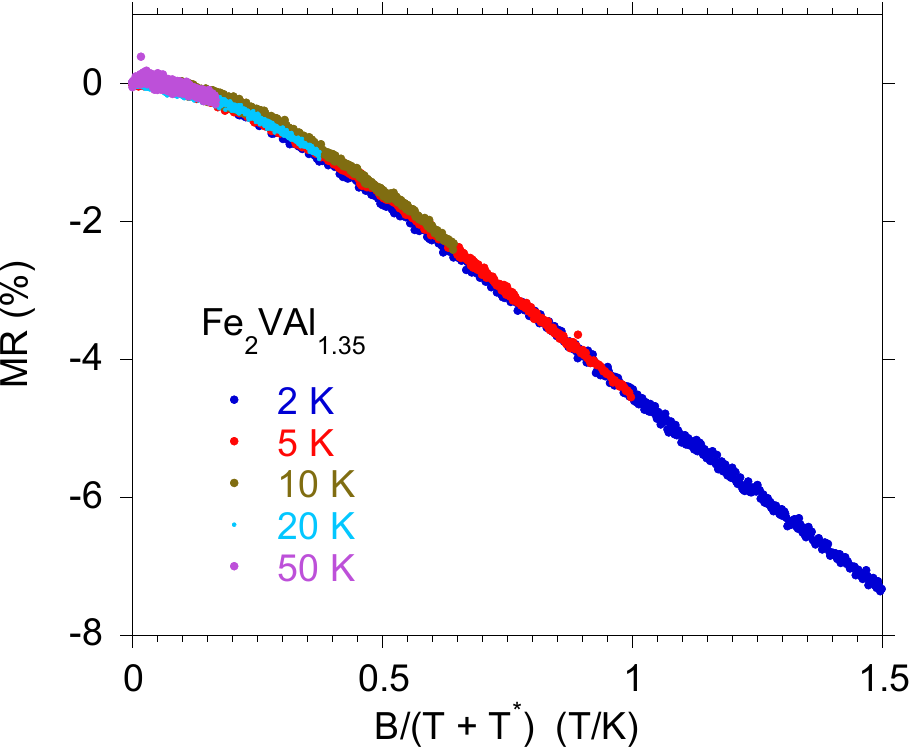}
\caption{\label{fig:MR_Fe_Skalowanie-Kondo}
Schlottmann-type plot of the magnetoresistance isotherms from panel Fig. \ref{fig:Fig_MR_Fe-Ni}(a). $T^{\star} = 2.9$ K.
}
\end{figure} 

Figure \ref{fig:Fig_MR_Fe-Ni}  compares magnetoresistance isotherms, $MR$,  of Fe$_2$VAl$_{1.35}$ (a) and Ni$_2$VAl$_{1.08}$ (b) measured from $0-9$ T. 
$MR$ is defined as
$MR= [\rho (B) - \rho (B = 0)]/\rho (B = 0)\times 100$\%, where $\rho (B)$  and $\rho (0)$ are resistivities measured at B=0  and H T, respectively. Since the applied magnetic field suppresses fluctuations in magnetic moments and spin-dependent scattering, a possible source of negative $MR$ can be Kondo behavior, as shown in Fig. \ref{fig:Fig_MR_Fe-Ni}(a).  The $MR$ isotherms  of Fe$_2$VAl$_{1.35}$ were found to be negative, supporting the Kondo effect as the significant  mechanism governing the low-$T$ electrical phenomena. Remarkably, as shown in Fig. \ref{fig:MR_Fe_Skalowanie-Kondo} the $MR$ isotherms taken
can be projected onto a single curve by plotting the
$MR$ data as a function of $B/(T + T ^{\star})$, where $T ^{\star}$ K is the characteristic temperature, usually considered as an approximate measure of the Kondo temperature \cite{Schlottmann1983}.
The Schlottmann-type scaling was applied to Fe$_2$VAl$_{1.35}$ giving $T ^{\star}=2.9$ K.
The $MR$ isotherms of Ni$_2$VAl$_{1.08}$ are quite different, $MR$ does not exhibit any field dependence as shown in Fig. \ref{fig:Fig_MR_Fe-Ni}(b) (cf. Fig. \ref{fig:Fig_Ni_R-inset}), even though Ni magnetic impurities contribute a term to the electrical resistivity that increases logarithmically on temperature as temperature $T$ is lowered. 
Maybe, the strongly diluted magnetic impurities give a weak effect, weaker than the spread of points on $MR$ isotherms.

\section{Effect of off-stoichiometry and site disorder on the electronic properties of F\lc{e}$_2$VA\lc{l}$_{1+\delta}$ within DFT calculations, comparison with stoichiometric F\lc{e}$_2$VA\lc{l} and N\lc{i}$_2$VA\lc{l}\label{sec:abinitio}} 

\begin{figure}[hbt!]
\includegraphics[width=0.45\textwidth]{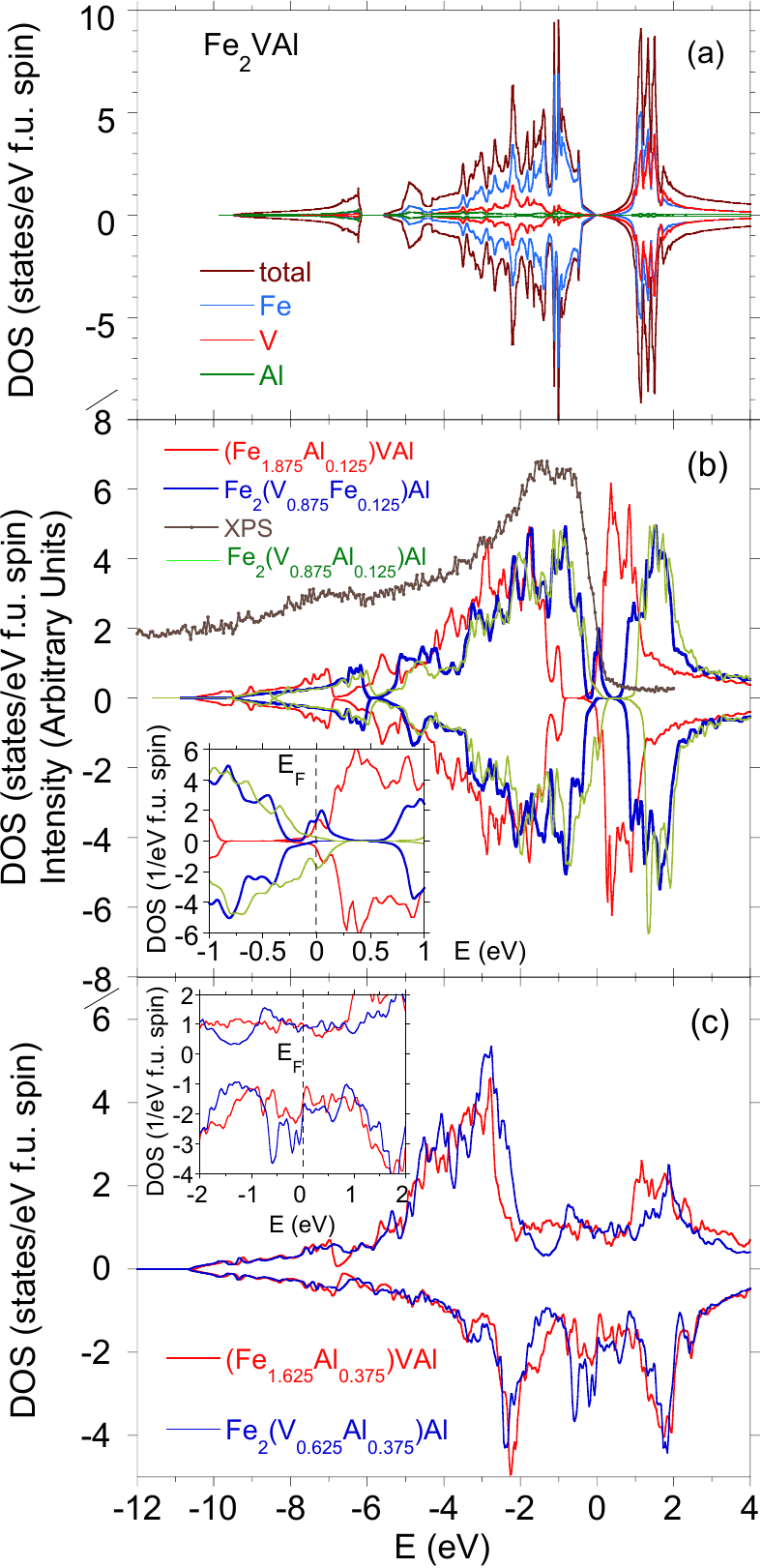}
\caption{\label{fig:Fig_DOS_XPS}
(a) Calculated total and spin-resolved density of states (TDOS) within the LSDA$+U+$SO ($U=3$ eV) approximation for Fe$_2$VAl. Also are shown the partial TDOS calculated for Fe, V, and Al.
(b) Valence band (VB) XPS spectrum for Fe$_2$VAl$_{1.35}$ (gray points) compared to the calculated TDOS (LSDA$+U+$SO, $U=3$ eV) for the supercell of various off-stoichiometry variants of Fe$_2$VAl. (c) The spin resolved  TDOS within the LSDA$+U+$SO, $U=3$ eV,  for (Fe$_{13}$Al$_{3}$)V$_8$Al$_8$ (red) and Fe$_{16}$(V$_{5}$Al$_3$)Al$_8$ (blue) supercells. The insets exhibit details near the Fermi level in an extended energy scale.
}
\end{figure}

\begin{figure}[hbt!]
\includegraphics[width=0.42\textwidth]{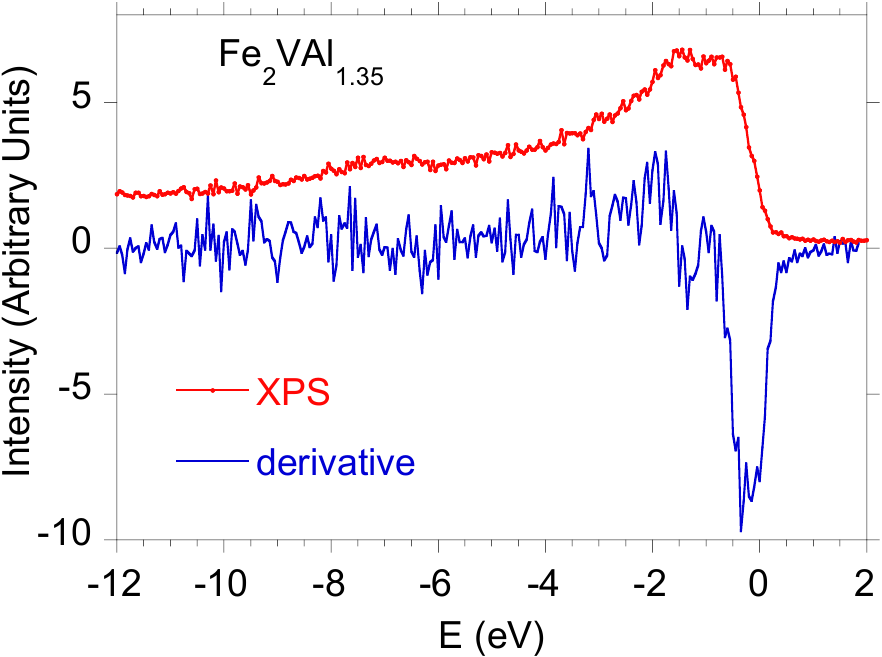}
\caption{\label{fig:Fig_Fe_XPS_deriv}
Valence band XPS spectrum for Fe$_2$VAl$_{1.35}$ (red points) and its derivative $dI/dE$ (blue line). The extreme in $dI/dE$ for $E\approx 0.2$ eV is related to Fe$_{AS}$ $d$-electron states located in the electronic bands near the Fermi level.  
}
\end{figure}

The electronic structure calculations for  Fe$_2$VAl  have shown that this compound is nonmagnetic and semi-metallic \cite{Singh1998}. The calculated density of states  of Fe$_2$VAl exhibits the 0.5 eV wide pseudogap located symmetrically around the Fermi level, as shown in Fig. \ref{fig:Fig_DOS_XPS}(a).
However, the electronic structure of this compound seems to be more interesting when Fe$_2$VAl is disordered or is off-stoichiometric. 
In the disordered Fe$_{2+x}$V$_{1-x}$Al composition  the Fe and V atoms at the antisite positions (F$\leftrightarrow$V) give rise to a narrow impurity $d$ band located just in the middle of the quasi-gap calculated for an ordered Fe$_2$VAl compound \cite{Deniszczyk2001}. This narrow $d$ band formed by the antisite Fe defects can significantly change the shape of the valence band XPS spectra of disordered alloy, especially near the Fermi level, as shown in Fig.\ref{fig:Fig_Fe_XPS_deriv}.
\begin{figure}[hbt!]
\includegraphics[width=0.45\textwidth]{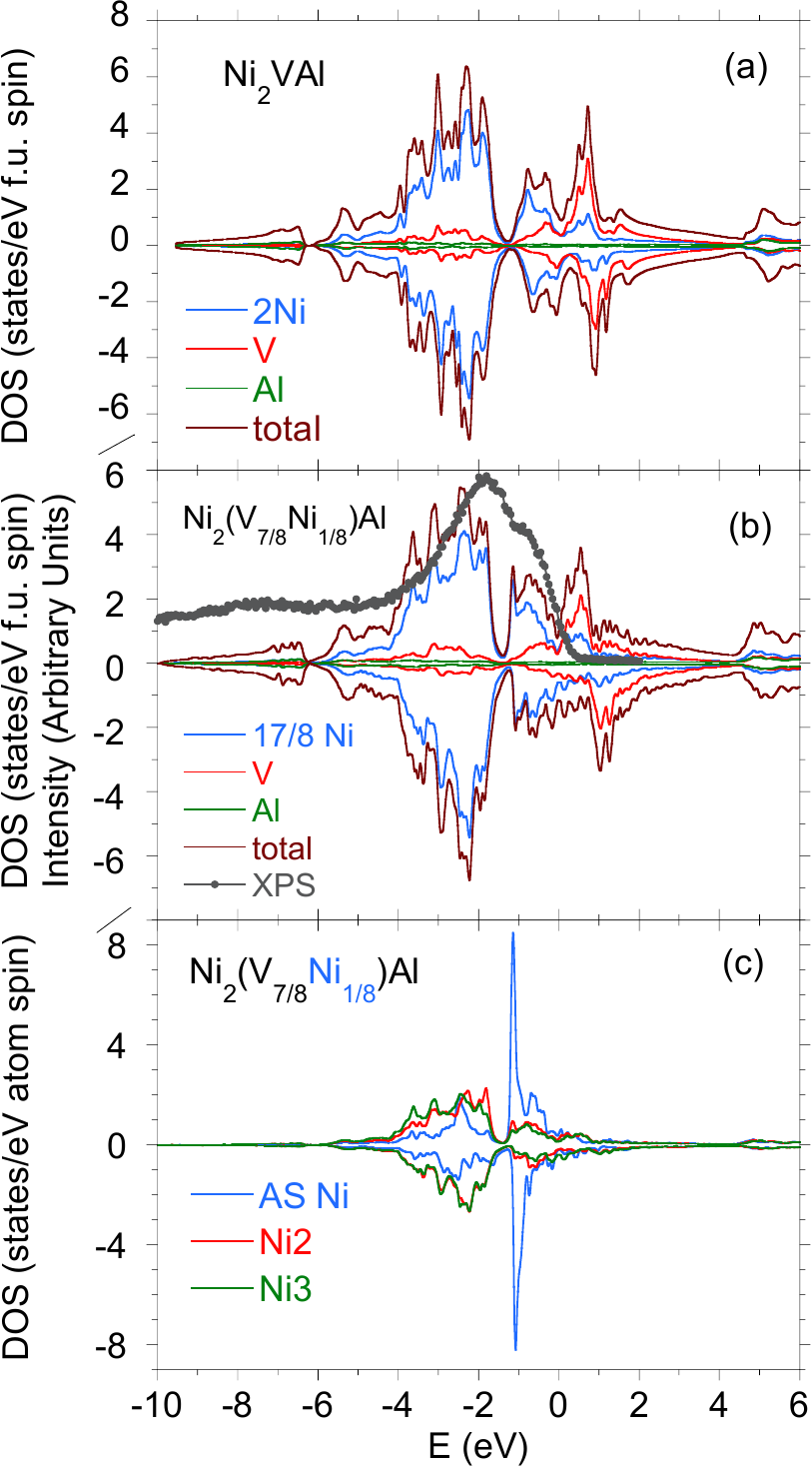}
\caption{\label{fig:Fig_XPS_DOS_Ni_A-B-C}
(a) Calculated total and spin resolved DOS within the LSDA$+U+$SO ($U=3$ eV) for Ni$_2$VAl. Also are shown the partial TDOS calculated for Ni, V, and Al.
(b) VB XPS spectrum for Ni$_2$VAl$_{1.08}$ (gray points) compared to the calculated TDOS (LSDA$+U+$SO, $U=3$ eV) for Ni$_{16}$(V$_{7}$Ni$_1$)Al$_8$ supercell. (c) Total DOS per one atom for Ni at various sites. The blue line represents the TDOS of AS Ni defects at V sites.
}
\end{figure}  
Appearance  of this  strongly correlated $d$-like band  was  also reported  for similar Fe$_2$TiSn compound due to an excess of Fe$_{AS}$ atoms at Ti antisite  positions \cite{Slebarski2006}. The DOS of this narrow peak in the gap of the Fe$_2$TiSn  bands is composed mainly of the $d-e_g$ states of Fe$_{AS}$ that hybridize with the $d$ states of the eight nearest Fe atoms in octahedral coordination; moreover, the calculated magnetic structure of [Fe$_{15}$Ti$_{AS}$][Ti$_{7}$Fe$_{AS}$]Sn$_{8}$ is of a cluster character (more details in Ref. \cite{Slebarski2004}).

The narrow $d$-band of  Fe$_{AS}$  is a reason of several anomalous thermodynamic properties attributed to many-body effects.
For example, the mechanism of electrical transport in Fe$_2$TiSn  with AS Fe defects has been explained as a result of interband transitions between this narrow $d$ band and other conduction states through a small gap at $\epsilon_F$ \cite{Slebarski2004}. 
Here, we calculated the bands for Fe$_2$VAl and Ni$_2$VAl in various variants of their stoichiometry and in the presence of the AS structural disorder. One notes,  Fe$_2$VAl$_{1.35}$ exhibits analogous $\rho(T)$ behavior to that of Fe$_2$TiSn, which suggests similar in nature electronic band properties of both the compounds.

The calculations were performed with the use of the FP-LAPW method complemented with local orbitals. The spin-orbit (SO) interaction
for valence and local orbital states was taken into account, and
the effective Hubbard parameter $U_{\rm eff}$ for the $d$-states of Fe, Ni, and V was
assumed to be 3 eV. The super-cell methodology of alloy modeling was used to simulate the disorder.
Figure \ref{fig:Fig_DOS_XPS} compares the atomic DOSs per atom for Fe$_2$VAl (a) with similar  DOSs calculated for supercell of the off-stoichiometric Fe$_2$VAl analogues containing an  excess of one Al atom located at Fe sites, (Fe$_{15}$Al$_{1}$)V$_8$Al$_8$, excess of one Fe at V sites, Fe$_{16}$(V$_{7}$Fe$_1$)Al$_8$, and  excess of Al at V sites, Fe$_{16}$(V$_{7}$Al$_1$)Al$_8$, respectively (b), and off-stoichiometric analogues shown in panel (c) where three Al atoms are at Fe sites, (Fe$_{13}$Al$_{3}$)V$_8$Al$_8$, and three Al atoms occupy  V sites, Fe$_{16}$(V$_{5}$Al$_3$)Al$_8$.
The first variant of disorder with one atom at the AS position predicts the gap located in the electronic bands of the assumed compounds; however, its location with respect to $\epsilon_F$ is different depending on the atom at the AS site. Namely, when an Al atom occupies the Fe sites, the gap is located at energies $E<\epsilon_F$, for one Fe at V sites the gap is symmetrically located around $\epsilon_F$, while for scenario with one Al at V sites the gap is located just above $\epsilon_F$. All scenarios give narrow band states at $\epsilon_F$, and are possible in the off-stoichiometry  sample.
However, note that only the AS Fe defects at V positions lead to the appearance of a narrow $3d\uparrow$ band at $\epsilon_F$, which seems to be the most reasonable explanation for the experimental data shown in Fig. \ref{fig:Fig_Fe_RvsT_H_inset}.
Within this scenario, the VRH hopping throw the pseudogap $k_BT_0$ of $\sim 20$ meV, which is calculated two orders in magnitude lower than the gap of Fe$_2$VAl,  can be  possible (cf. Fig. \ref{fig:Fig_DOS_XPS}(b)). It is reasonable to assume that all three DOSs shown in panel (b) contribute to the total DOS of the disordered sample with appropriate weighting factors, with the result giving an agreement of calculations with the valence XPS spectra. The DOSs calculated for the more disordered system, with the exchange of three Al atoms with Fe or V assumed, respectively, do not give a good comparison with the experiment, as shown in panel (c).
The details of the DFT calculations are summarized in Table \ref{tab:TableDFT}.
\begin{table*}[hbt!]
\caption{Fe$_2$VAl (F$m\bar{3}m$) and the off-stoichiometric analogues; FP-LAPW results for supercell of (Fe$_{15}$Al$_{1}$)V$_8$Al$_8$, Fe$_{16}$(V$_{7}$Fe$_1$)Al$_8$, and Fe$_{16}$(V$_{7}$Al$_1$)Al$_8$. $n_l$ is a number of valence electrons, $l=0$, 1, 2, and 3 ($s$, $p$, $d$, and $f$ electron states, respectively). $m$ is total magnetic moment of each atom in the supercell.
}
\label{tab:TableDFT}
\begin{tabular}{ccccccc}
\hline\hline
supercell   & formula unit  &  m$_{total}$  &   DOS($\epsilon_F$) & $\gamma_0^{calc}$  & &\\   
            &               & ($\mu_B$/f.u.) &   (1/eV f.u.)  & (mJ/K$^2$ mol)      & & \\   
\hline
            &  Fe$_2$VAl    &   0           &      0              &  0                 && \\
atom & multiplicity & $n_s$ & $n_p$ & $n_d$ & $n_f$ & m$_{total}$/atom in $\mu_B$  \\ 
\hline
Al  &   1   &  2.4354  &  6.5068 & 0.1094  & 0.0120  &  0.000      \\ 
Fe  &   2   &  2.2332  &  6.1952 & 5.8200  & 0.0070  &  0.000      \\
V   &   1   &  2.1458  &  6.0268 & 2.4886  & 0.0152  &  0.000      \\
\hline
Fe$_{15}$V$_8$Al$_9$   &  (Fe$_{15/8}$Al$_{1/8}$)VAl  &   0.04           &      1.93      &  4.55         && \\
atom & multiplicity & $n_s$ & $n_p$ & $n_d$ & $n_f$ & m$_{total}$/atom in $\mu_B$  \\ 
\hline
Fe1&  1 & 2.24 & 6.20 & 5.83 & 0.01 &  0.155 \\
Fe2&  1 & 2.24 & 6.20 & 5.83 & 0.01  & 0.082 \\
Fe3&  1 & 2.23 & 6.19 & 5.82 & 0.01  &-0.123 \\
Fe4&  4 & 2.23 & 6.19 & 5.82 & 0.01  &-0.052 \\
Fe5&  2 & 2.23 & 6.19 & 5.82 & 0.01  &-0.056 \\
Fe6&  4 & 2.23 & 6.19 & 5.81 & 0.01  & 0.012 \\
Fe7&  2 & 2.23 & 6.19 & 5.81 & 0.01  & 0.018 \\
V1&   4 & 2.15  &6.03 & 2.46 & 0.01  & 0.056 \\
V2&   4 & 2.15  &6.03 & 2.42 & 0.01  & 0.039 \\
Al1&  4 & 2.43 & 6.50 & 0.12 & 0.01  & 0.000 \\
Al2&  4 & 2.43 & 6.50 & 0.11 & 0.01  & 0.000 \\
Al3&  1 & 2.43 & 6.52 & 0.10 & 0.01  & 0.000 \\
\hline
Fe$_{16}$V$_7$Al$_9$   &  Fe$_{2}$(V$_{7/8}$Al$_{1/8}$)Al  &   0.21           &     2.18     &  5.13        && \\
atom & multiplicity & $n_s$ & $n_p$ & $n_d$ & $n_f$ & m$_{total}$/atom in $\mu_B$  \\ 
\hline
Al1&  2 & 2.4350& 6.5119& 0.1190& 0.0121& -0.00574\\
Al2&  4 & 2.4330& 6.5040& 0.1147& 0.0117& -0.00542\\
Al3&  2 & 2.4329& 6.5037& 0.1147& 0.0118& -0.00509\\
Fe1&  8 & 2.2284& 6.1911& 5.8335& 0.0065&  0.34381\\
Fe2&  8 & 2.2332& 6.1961& 5.8190& 0.0070&  0.01825\\
Al4&  1 & 2.4260& 6.4748& 0.1119& 0.0111& -0.02047\\
V1&  1  &2.1520& 6.0366 &2.4217& 0.0157 & 0.00161\\
V2&   4& 2.1508& 6.0356& 2.4231& 0.0154& -0.16059\\
V3&   2 & 2.1508& 6.0357& 2.4225& 0.0154& -0.15168\\
\hline
Fe$_{17}$V$_7$Al$_8$   &  Fe$_{2}$(V$_{7/8}$Fe$_{1/8}$)Al  &   0.38          &     1.078     &  2.54        && \\
atom & multiplicity & $n_s$ & $n_p$ & $n_d$ & $n_f$ & m$_{total}$/atom in $\mu_B$  \\ 
\hline
Fe1&  1 & 2.2377 &6.1858& 5.5736& 0.0117 & 2.98449\\
Fe2&  8 & 2.2261& 6.1824& 5.8144& 0.0065&  0.46897\\
Fe3&  8 & 2.2292 &6.1881& 5.8108& 0.0066& -0.41870\\
V1&   4 & 2.1479& 6.0294& 2.4138& 0.0142& -0.05980\\
V2&   2 & 2.1479& 6.0299& 2.4149& 0.0143& -0.04267\\
V3&   1&  2.1482& 6.0304& 2.4103& 0.0145&  0.29083\\
Al1&  2 & 2.4315& 6.4919& 0.1079& 0.0110&-0.00698\\
Al2&  4 & 2.4317& 6.4923& 0.1086& 0.0110& -0.00697\\
Al3&  2& 2.4300& 6.4984& 0.1122& 0.0110& -0.00303\\
\hline\hline
\end{tabular} 
\end{table*}

\begin{table*}[hbt!]
\caption{Ni$_2$VAl (F$m\bar{3}m$) and the off-stoichiometric analogues; FP-LAPW results for supercell of 
Ni$_{16}$(V$_{7}$Ni$_1$)Al$_8$. $n_l$ is a number of valence electrons, $l=0$, 1, 2, and 3 ($s$, $p$, $d$, and $f$ electron states, respectively). $m$ is total magnetic moment of each atom in the supercell.}
\label{tab:DOS_Ni}
\begin{tabular}{ccccccc}
\hline\hline
supercell   & formula unit  &  m$_{total}$  &   DOS($\epsilon_F$) & $\gamma_0^{calc}$  & &\\   
            &               & ($\mu_B$/f.u.) &   (1/eV f.u.)  & (mJ/K$^2$ mol)      & & \\   
\hline
            &  Ni$_2$VAl    &  0.335         &     3.06       &  7.22       && \\
atom & multiplicity & $n_s$ & $n_p$ & $n_d$ & $n_f$ & m$_{total}$/atom in $\mu_B$ \\ 
\hline
Ni  &   2   &  2.2946  &  6.2264 & 7.9708  & 0.0051  &  0.00649      \\
V   &   1   &  2.1273  & 6.0018 &  2.4359  & 0.0103  & 0.29991     \\
Al  &   1   &  2.4171  & 6.4704 &  0.0835  & 0.0084  & -0.00541    \\ 
\hline
Ni$_{17}$V$_7$Al$_8$   &  Ni$_{2}$(V$_{7/8}$Ni$_{1/8}$)Al  &   0.38          &     2.361     & 5.57      && \\
atom & multiplicity & $n_s$ & $n_p$ & $n_d$ & $n_f$ & m$_{total}$/atom in $\mu_B$  \\ 
\hline
Ni1&  1 & 2.2244 &6.1632 &7.9765 &0.0060  &-0.01783\\
Ni2&  8 & 2.2887 &6.2181 &7.9810 &0.0055  &-0.06997 \\
Ni3&  8 & 2.2896 &6.2269 &7.9605 &0.0054  & 0.14194\\
V1&   4 &2.1293 &6.0035 &2.4393 &0.0109   &0.57564\\
V2&   2 &  2.1293 &6.0035 &2.4417 &0.0108 &  0.61413\\
V3&   1 & 2.1294 &5.9973 &2.4440 &0.0108  & 0.87389\\
Al1&  2 & 2.4157 &6.4562 &0.0813 &0.0083  &-0.00698 \\
Al2&  4 & 2.4157 &6.4564 &0.0811 &0.0083  &-0.00753\\
Al3&  2 & 2.4180 &6.4798 &0.0921 &0.0093  &-0.00782 \\
\hline\hline
\end{tabular} 
\end{table*}
Fig. \ref{fig:Fig_XPS_DOS_Ni_A-B-C} shows the valence band  XPS spectra in comparison to the calculated total DOSs per formula for Ni$_2$VAl (a) and Ni$_{17}$V$_7$Al$_8$ supercell (b). This comparison clearly shows that the measured VB XPS spectra mostly reflect the $3d$ Ni and  $3d$ V electronic states located between the Fermi energy and the binding energy $6$ eV. 
The Al $3ps$ states are located between 6 and 10 eV below $\epsilon_F$. One can note that the Ni AS defects do not drastically change the total DOSs of Ni$_2$VAl, however, they significantly contribute to the sharp $d$-electronic states at 1 eV below $\epsilon_F$ [in panel (c)], giving a magnetic moment of $0.14$ $\mu_B$ on Ni at AS positions (cf. Table \ref{tab:DOS_Ni}). 
A localized magnetic moment calculated for AS Ni correlates well with the $\rho\sim -\ln T$ behavior, characteristic of the diluted Kondo systems (as shown in Fig. \ref{fig:Fig_Ni_R-inset}).
\begin{figure}[hbt!]
\includegraphics[width=0.45\textwidth]{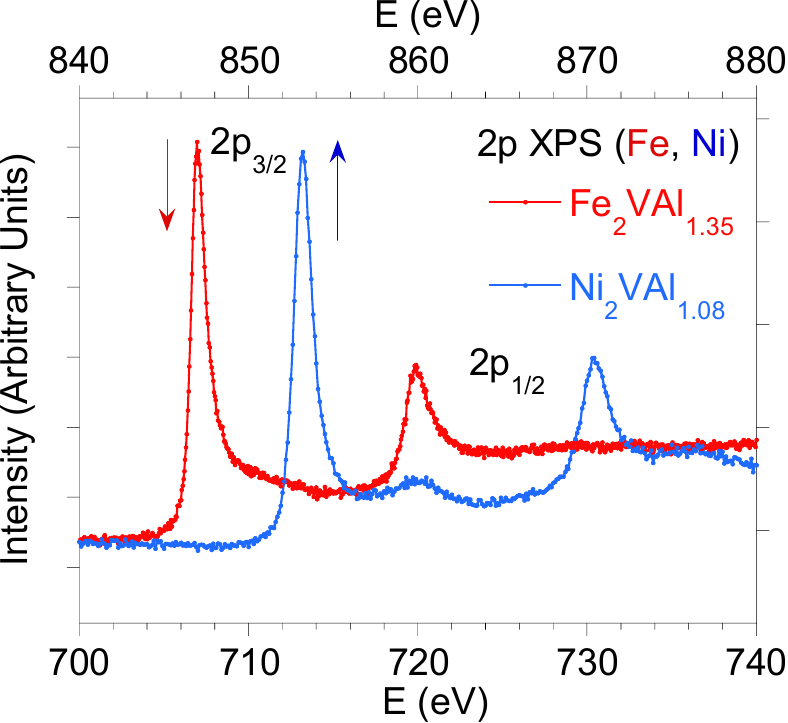}
\caption{\label{fig:XPA_Fe-Ni_2p_sum}
$2p$ XPS core level spectra for Fe in Fe$_2$VAl$_{1.35}$  (red) and Ni in 
Ni$_2$VAl$_{1.08}$ (blue).
}
\end{figure} 
Finally, we have measured the core level $2p$ XPS spectra for Fe in Fe$_2$VAl$_{1.35}$ and Ni in Ni$_2$VAl$_{1.08}$ to indirectly argument the low avarage magnetic moment of Fe and Ni in these compounds. The $2p$ XPS spectra are interpreted in reference to Refs. \cite{Neumann1,Neumann2}, where for Mn-based Heusler alloys, it has been documented that the exchange splitting of the $2p_{3/2}$ level is directly correlated with the value of local magnetic moment at the Mn site. Specifically, the $2p_{3/2}$ splitting energy $\Delta E$ plotted as $\Delta E$  versus magnetic moment $\mu_{Mn}$ of Mn for a series of various Mn-based Heusler alloys exhibits a universal linear dependence, which gives for $\mu_{Mn}\sim 3$ $\mu_B$ a value of $\Delta E\sim 1$ eV. This experimental observation seems to be universal, and characteristic of other $d$-electron  elements with localized magnetic moment larger than 2 $\mu_B$/atom (cf. Ref. \cite{Slebarski2001}), simultaneously the XP spectroscopy allows one to quickly demonstrate the strength of $\mu$.
The $2p_{3/2}$ lines shown in Fig. \ref{fig:XPA_Fe-Ni_2p_sum} do not exhibit any $\Delta E$ splitting, which suggests a nonmagnetic ground state or strongly delocalized $d$-electronic states or both, or that the magnetic moment localized on Fe and Ni, respectively,  is not large enough to observe the splitting.

\section{Concluding remarks}

Spin fluctuations in itinerant electron systems could give predominant effects on the thermodynamic properties of wealky or nearly ferromagnetic metals \cite{Moriya1985}. In consequence of the appearance of spin fluctuations, one expects an enhance of electronic specific heat, Pauli susceptibility, as well as a strong enhance of the thermopower. For many reasons, the disordered Fe$_2$VAl has been a candidate for good thermoelectric properties. In several previously published papers, the thermoelectric properties have been studied, either experimentally and theoretically \cite{Tsujii2019,Naydenov2019,Bourgault2023} for pure Fe$_2$VAl and with dopants, however, the effect did not meet expectations. We investigated the off-stoichiometric Fe$_2$VAl$_{1.35}$ with excess of Al, however, the sample is not appropriate for thermoelectric applications. However, the sample is extremely interesting because of the complex magnetic properties caused by disorder. We documented the impact of AS disorder on the appearance of magnetic moment on Fe in AS positions, which in consequence leads to appearance of Griffiths phase state below $T_G\approx 200$ K, and singular properties in low-temperature susceptibility. In the paramagnetic regime inverse susceptibility  of Fe$_2$VAl$_{1.35}$ obeys the Curie-Weiss law, while below a characteristic temperature $T_G$  $1/\chi$ displays a downward deviation from the CW law with evidently field dependent behavior,  indicating the onset of short-range ferromagnetic correlation well above $T_C^g$, which is considered a hallmark of Griffiths singularity. 

The DFT calculations  were carried out for the disordered Fe$_2$VAl and its off-stoichiometric variants. The {\it ab initio} calculations predicted the  Fe$_2$VAl compound to be 
{\it nonmagnetic} narrow-gap semiconductor, while for similar disordered and/or off-stoichiometric alloys  Fe occupying the AS V sites is always calculated magnetic. 
Both band structure calculations and magnetic measurements showed at most one AS Fe defect per unit cell; therefore, the system can be treated as dilute. In result, the Griffiths phase state is possible in such a diluted system due to the finite probability of randomly large, pure, and differently diluted clusters.   This result allowed us to simulate the magnetization versus temperature within the Ising model in
an external magnetic field, in good agreement with the experimental data shown in Fig. \ref{fig:Fig_CHI_Fe_sum_inset}. 

The physical properties of Fe$_2$VAl$_{1.35}$ are analyzed with respect to paramagnetic Ni$_2$VAl. Up to now, this compound has not been sufficiently well investigated, more of its possible behaviors have been predicted from DFT calculations \cite{Rocha1999,Reddy2016,Wen2017,Wang2020}. We present comprehensive thermodynamic investigations as well as transport properties for this compound. Our DFT calculations predict magnetic moment on Ni at AS V sites, which is a reason of appearance of  Kondo diluted effect in the low-temperature resistivity of Ni$_2$VAl, however, the superconductivity demonstrated by {\it ab initio} calculations, as was reported by Sreenivasa {\it et al.} \cite{Reddy2016},  has not been confirmed experimentally for this compound.

\appendix
\section{Numerical analysis}

\begin{figure}[h!]
\includegraphics[width=0.48\textwidth]{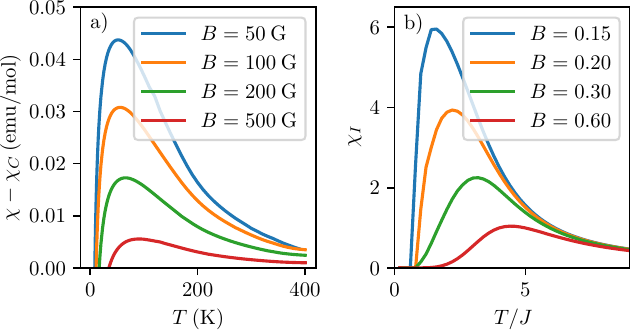}
\caption{a) Difference between the measured susceptibility $\chi(T)$ and susceptibility given by the Curie law $\chi_C$. Only part for $\chi > \chi_C$ is shown. b) Magnetic susceptibility $\chi_I$ of a $2\times 2\times 2$ Ising model in external magnetic field. 
\label{fig:clusters}}
\end{figure}

In Sec.~\ref{sec:magn_prop} we argue that the deviation from the CW law is driven by the formation of ferromagnetic clusters. As discussed in Sec.~\ref{sec:abinitio}, {\it ab initio} calculations indicate the presence of magnetic moments on wrong-site iron atoms. The divergence of susceptibility $\chi$ (cf. Fig.~\ref{fig:CHI_dc_Fe-Ni_inset}) suggests a ferromagnetic interaction between them. However, since down to the lowest temperatures studied, the system is in the paramagnetic state, those moments are too diluted to develop a ferromagnetic state. Moreover, as can be seen in Fig.~\ref{fig:Fig_CHI_Fe_sum_inset}a), $\chi(T)$ clearly deviates from the Curie law $\chi_C\propto T^{-1}$, which we attribute to magnetic clusters formed in the Griffiths phase. To support this assumption, we performed Monte Carlo simulations for small magnetic clusters to determine the temperature dependence of their contribution to the bulk magnetic susceptibility. We assume that the main signal comes from the bulk of the system and fulfills the Curie law, so determine the contribution from magnetic clusters, in Fig.~\ref{fig:clusters}a) we present the difference between the measured susceptibility $\chi$ and $\chi_C$ for different values of the magnetic field.

In Fig.~\ref{fig:clusters}b), for comparison, we present a temperature dependence of the magnetic susceptibility of a very small ($2\times 2\times 2$) cluster described by the Ising model in an external magnetic field. However, there are two differences between these two results. The first is related to the behavior at low temperatures. On the one hand, the measured susceptibility $\chi$ remains finite when $T\to 0$, while $\chi_C$ diverges as $T^{-1}$. Therefore, $\chi-\chi_C$ goes to $-\infty$ when $T\to 0$ (only $\chi > \chi_C$ is shown in Fig.~\ref{fig:clusters}a). On the other hand, the susceptibility $\chi_I$ of the Ising model goes to zero when $T\to 0$. This discrepancy can result from approximations/limitations applied to $\chi-\chi_C$, e.g., instead of the Curie law, the bulk susceptibility can be described by the Curie-Weiss law with $\theta_{\rm CW}<0$ which does not have a singularity at $T=0$. The other difference between panels a) and b) of Fig.~\ref{fig:clusters} can be seen at high temperature, where $\chi_I$ is almost temperature independent, while $\chi-\chi_C$ decreases with increasing temperature. This discrepancy can result from neglecting other contributions to the magnetic susceptibility, e.g., the effect of spin fluctuations. The qualitative differences between the shapes of the lines are also due to the vast simplification of considering only one size of the clusters. In the Griffiths phase we expect an ensemble of clusters of different sizes with their distribution depending on temperature. Moreover, we took into account the simplest possible kind of magnetic interactions in the cluster, i.e, the Ising model. This simplification does not allow for different magnetization orientations in different clusters, which is expected because of the complex nature of the RKKY interaction. However, the qualitative agreement of the temperature dependencies of $\chi-\chi_C$ and $\chi_I$ we observe despite the use of strong approximations applied both to $\chi-\chi_C$ and to the modeled $\chi_I$ is a strong argument for the presence of magnetic clusters in the system.

$^{\star}$Author to whom correspondence should be addressed: andrzej.slebarski@us.edu.pl

\begin{acknowledgments}
Numerical calculations have been carried out using High Performance Computing resources provided by the Wrocław Centre for Networking and Supercomputing.
\end{acknowledgments}


\begin{thebibliography}{50}
\bibitem{comment3}
$n=0$ corresponds to Debye-type exponential relaxation, for $n=1$, there is no relaxation at all.
\end{thebibliography}
\end{document}